\documentclass[12pt]{article}
\pdfoutput=1
\usepackage[utf8]{inputenc}
\usepackage{dsfont}
\usepackage{amsfonts}
\usepackage{amsmath}
\allowdisplaybreaks[4]        
\usepackage{starfont}
\usepackage{amssymb}
\usepackage{euscript}     
\usepackage{braket}
\usepackage{color,soul}         
\usepackage{tensor}        
\usepackage{amsthm}
\usepackage{graphicx}
\usepackage{slashed}
\usepackage{subfigure}
\usepackage{bbm}
\definecolor{outerspace}{rgb}{0.25, 0.29, 0.3}
\definecolor{scarlet}{rgb}{1.0, 0.13, 0.0}
\usepackage[header,title,page,titletoc]{appendix}  
\definecolor{princetonorange}{rgb}{1.0, 0.56, 0.0}
\definecolor{WildStrawberry}{rgb}{1.0, 0.26, 0.64}
\definecolor{rossocorsa}{rgb}{0.83, 0.0, 0.0}
\definecolor{navyblue}{rgb}{0.0, 0.0, 0.5}
\usepackage[numbers,sort&compress]{natbib}  
\usepackage{float}
\usepackage[paper=letterpaper,margin=1in]{geometry}
\newcommand{\req}[1]{(\ref{#1})} 

\newcommand{\bea}{\begin{eqnarray}}
\newcommand{\eea}{\end{eqnarray}}
\newcommand{\ba}{\begin{eqnarray}}
\newcommand{\ea}{\end{eqnarray}}

\newcommand{\beq}{\begin{equation}}
\newcommand{\eeq}{\end{equation} }
\newcommand{\beqa}{\begin{eqnarray}}
\newcommand{\eeqa}{\end{eqnarray}}
\newcommand{\beqar}{\begin{eqnarray*}}
\newcommand{\eeqar}{\end{eqnarray*}}

\renewcommand{\req}[1]{eq.~(\ref{#1})}

\newcommand{\eg}{{\it e.g.,}\ }
\newcommand{\ie}{{\it i.e.,}\ }

\newcommand{\cH}{\mathcal{H}}
\newcommand{\cL}{\mathcal{L}}

\newcommand{\cO}{\mathcal{O}}
\newcommand{\cM}{\mathcal{M}}

\usepackage{hyperref}
\hypersetup{
    colorlinks,
    citecolor=rossocorsa,
    filecolor=navyblue,
    linkcolor=navyblue,
    urlcolor=navyblue
}

\begin{document} 

\begin{titlepage}

\begin{center}

\phantom{ }
\vspace{3cm}

{\bf \Large{Complexity measures in QFT\\ and constrained geometric actions}}
\vskip 0.5cm
Pablo Bueno,${}^{\text{\Zeus}}$ Javier M. Mag\'an${}^{\text{\Zeus}}$ and C. S. Shahbazi${}^{\text{\Kronos}}$
\vskip 0.05in
\small{$^{\text{\Zeus}}$ \textit{Instituto Balseiro, Centro At\'omico Bariloche}}
\vskip -.4cm
\small{\textit{ 8400-S.C. de Bariloche, R\'io Negro, Argentina}}

\small{${}^{\text{\Kronos}}$ \textit{Department of Mathematics, University of Hamburg}}
\vskip -.4cm
\small{\textit{ Bundesstra\ss e 55, 20146, Hamburg, Germany}}

\begin{abstract}
We study the conditions under which, given a generic quantum system, complexity metrics provide actual lower bounds to the circuit complexity associated to a set of quantum gates. Inhomogeneous cost functions ---many examples of which have been recently proposed in the literature--- are ruled out by our analysis. Such measures are shown to be unrelated to circuit complexity in general and to produce severe violations of Lloyd's bound in simple situations.
Among the metrics which do provide lower bounds, the idea is to select those which produce the tightest possible ones. This establishes a hierarchy of cost functions and considerably reduces the list of candidate complexity measures. In particular, the criterion suggests a canonical way of dealing with penalties, consisting in assigning infinite costs to directions not belonging to the gate set. We discuss how this can be implemented through the use of Lagrange multipliers. We argue that one of the surviving cost functions defines a particularly canonical notion in the sense that: i) it straightforwardly follows from the standard Hermitian metric in Hilbert space;  ii) its associated complexity functional is closely related to Kirillov's coadjoint orbit action, providing an explicit realization of the ``complexity equals action'' idea; iii) it arises from a Hamilton-Jacobi analysis of the ``quantum action'' describing quantum dynamics in the phase space canonically associated to every Hilbert space. Finally, we explain how these structures provide a natural framework for characterizing chaos in classical and quantum systems on an equal footing, find the minimal geodesic connecting two nearby trajectories, and describe how complexity measures are sensitive to Lyapunov exponents. 
\end{abstract}
\end{center}
\end{titlepage}

\setcounter{tocdepth}{2}

{\parskip = .2\baselineskip \tableofcontents}

\section{Introduction}
\label{sec:Introduction}
Triggered by Susskind \emph{et al.}'s observation that the ``size''\footnote{The most popular notions of ``size'' conjectured to be connected to quantum complexity are the ``complexity equals volume'' and ``complexity equals action'' proposals. They conjecture, respectively, that the complexity of the CFT state is related to the volume of the extremal codimension-one bulk region meeting the boundary on the corresponding time slice \cite{Susskind:2014rva,Stanford:2014jda,Susskind:2014jwa,Susskind:2014moa} and that it is related to the gravitational action evaluated on the domain of dependence of bulk Cauchy slices asymptotically approaching such boundary time slice \cite{Brown:2015bva,Brown:2015lvg}. See also \cite{Couch:2016exn} for an alternative proposal. } of black hole interiors grows with time in a remarkably similar fashion to the computational complexity of discrete quantum systems, and making crucial use of  Nielsen \emph{et al.}'s geometric approach \cite{2005quant.ph..2070N,2006Sci...311.1133N,2007quant.ph..1004D}, numerous attempts at defining reasonable notions of complexity in the context of quantum field theory (QFT) have been explored in recent times.

This task is challenging for various reasons. Roughly speaking ---see below for a more precise definition---  the complexity of a given unitary transformation is defined as the smallest number of ``small'' unitaries, or ``gates'', belonging to certain universal set required for implementing such transformation. This setup is particularly well-suited for $n$-qubit systems. In that context, ``big'' unitaries correspond to generic $n$-site tensor products of single-qubit Pauli matrices and identities, and a canonical set of ``small'' gates is given by $1$- and $2$-qubit gates, which suffice to generate any big unitary. The situation is not as easy in the case of continuous systems, for which notions like ``big unitaries'' or ``gates'' become considerably more obscure. An interesting development ---which preceded all existent literature on QFT complexity--- was carried out by Nielsen and collaborators \cite{2005quant.ph..2070N,2006Sci...311.1133N,2007quant.ph..1004D}, who showed that the complexity of $n$-qubit systems can be approximated by the length of geodesics ---associated to certain notions of distance--- on the unitaries manifold. In particular, they proved that, whenever such distance functionals satisfied certain conditions,  they were able to provide lower bounds on the circuit complexity. 

This ``continuous approach'' appears to be suitable for quantum fields and has indeed been at the root of most ---albeit not all--- proposals discussed so far in the literature.
One of the first involved fidelity susceptibility ---a quantity which is equivalent to the so-called quantum information metric--- in the context of small perturbations of the thermofield double state \cite{MIyaji:2015mia}. Subsequent ideas involved applying a Nielsen-inspired setup to free QFTs \cite{Jefferson:2017sdb,Chapman:2017rqy}. These and related proposals have been subsequently explored in many papers ---see \eg \cite{Yang:2017nfn,Khan:2018rzm,Hackl:2018ptj,Camargo:2018eof,Guo:2018kzl,Bhattacharyya:2018bbv,Chapman:2018hou,Jiang:2018nzg,Jiang:2018gft,Akal:2019ynl,Bhattacharyya:2019kvj,Liu:2019aji,Camilo2019}. Exploiting the power of restricting the gate set to a Lie subgroup of the unitaries manifold, a proposal valid for generic (interacting) CFTs was presented in \cite{Magan:2018nmu}, and further developed in \cite{Caputa:2018kdj}, where it was connected with the field of ``coadjoint orbit actions'' \cite{Kiri}. Somewhat related to this approach, Euler-Arnorld actions/equations have also been suggested as complexity measures \cite{Caputa:2018kdj,Balasubramanian:2018hsu}. Fubini-Study and fidelity susceptibility-like proposals were further explored in \cite{Belin:2018bpg} for interacting CFTs. Yet another approach, known as ``path integral complexity'', was proposed in \cite{Caputa:2017urj,Caputa2017}. This  relates complexity to the variation of the path integral measure under conformal transformations.\footnote{This proposal was explored in the context of circuit complexity in \cite{Caputa:2018kdj}. See also \cite{Camargo:2019isp}.}

Of course, the existence of this zoo of proposals rises up many questions. In this paper we focus on three obvious ones: i) how do we know a given   proposal provides a reasonable complexity measure?;  ii) are there any guiding principles we can use to establish some kind of hierarchy among the alternatives?;  iii) what is the relation between complexity measures/actions and the actual physical actions describing the dynamics of a quantum system?

While it is tempting to use continuous notions directly as definitions for complexity, Nielsen's original observation that only certain notions provide lower bounds for that quantity suggests in fact a way to address the first two questions.  Given a complexity metric, we have to make sure that for any discrete protocol, generated by a discrete set of quantum gates, it is always possible to construct a continuous one such that its length, as computed by the complexity metric, is smaller  than (or equal to) the number of gates used in the discrete protocol. Metrics satisfying this condition were called $\mathcal{G}$-bounding in \cite{2005quant.ph..2070N} in the context of qubit systems. We will see that an analogous connection between $\mathcal{G}$-bounding metrics and lower bounds to complexity applies to generic quantum systems, providing a precise criterion for determining whether a proposed measure is actually related to complexity. Interestingly, this simple analysis already rules out inhomogeneous metrics, examples of which have been proposed independently by several groups. Among all possible metrics satisfying the $\mathcal{G}$-bounding condition, those attributing greater distances to generic protocols should be preferred over those attributing smaller distances. The philosophy is simple: given a $\mathcal{G}$-bounding metric, the tighter the lower bound it provides, the closer it gets to the actual complexity of a given protocol. We will see how this principle constrains the zoo of complexity metrics as well as the so-called penalty functions.

The outcome of the analysis reveals that the most natural tight bound is provided by the standard metric induced by the usual Hermitian product in Hilbert space, plus constraints implementing infinite penalties for directions falling outside the gate set. The corresponding cost function is denoted $F_{\braket{H^2}}$ throughout the text and it appears defined in \req{H2Ja} below.
This standard metric in Hilbert space is associated to a canonical symplectic structure which allows one to understand the usual quantum evolution as a (classical) Hamiltonian evolution ---see \eg \cite{Kibble:1978tm,Ashtekar:1997ud} for standard references on this subject. This fact, along with the Lie group structure of the unitaries manifold gives rise to a canonical distance functional ---the so-called ``coadjoint orbit'' or ``geometric'' action--- on the space of unitaries. Interestingly, such a notion is intimately related to the aforementioned complexity measure  $F_{\braket{H^2}}$ ---they are equal for systems of small quantum variance. Hence, besides giving rise to tight lower bounds for circuit complexity, $F_{\braket{H^2}}$ provides a realization of the ``complexity equals action'' idea, where ``action'' stands here for the geometric action canonically associated to the system.\footnote{For 2d CFTs, such action is equivalent to Poliakov's two dimensional gravity when pulled back to the coadjoint orbits of the Virasoro group \cite{Alekseev:1988ce}. This was realized in \cite{Caputa2017} from a quantum complexity perspective using the cost $F_{\braket{H^2}}$.}

With regards to the third question, we use the canonical symplectic structure associated to any Hilbert space to construct a ``quantum action'' (given a quantum Hamiltonian), whose classical dynamics is just the Schr\"odinger equation. This quantum action may be viewed at the same level as the complexity actions, in the sense that it is a classical action defined on Hilbert space. The advantage is that we know this quantum action to be the one controlling the exact quantum dynamics of the system (in a classical manner).\footnote{A further nice feature of this action is that it can be shown to reduce to the right classical actions in semiclassical approximations.} Using this quantum action, we show how the integrand of the geometric action arises from the Hamilton-Jacobi equation associated to it, and in this precise sense it is related to the actual dynamics of a given system.

Finally, the formulation of quantum mechanics as a classical system in the Hilbert space also allows to clarify the relation between complexity and chaos. We study this in the final section, where we propose to use the conventional classical definition of chaos, but now in the quantum phase space ---\ie the Hilbert space. This definition has the right pullback to any semiclassical phase space immersed within the Hilbert space, but it is otherwise valid for the full quantum system. Using simple intuition from classical physics, the minimal geodesic connecting two nearby trajectories can now be found, and we can frame the chaotic process as a simple quantum circuit. In particular, we relate the instantaneous Hamiltonian to the so-called Jacobian matrix controlling the linearized classical dynamics, from which Lyapunov exponents can be computed. We discuss how different measures are sensitive (or not) to the expected exponential growth.

The paper is organized as follows. In Section \ref{sec:BLgroups} we comment on the mathematical structure of the unitaries group of a generic quantum system when this is infinite-dimensional.
In Section \ref{bounds} we show how Nielsen's proof of the connection between the $\mathcal{G}$-bounding condition and lower bounds on complexity naturally extends to any quantum system, including QFTs. We start with a review of the definition, difficulties and technicalities associated to continuous protocols. We then review the zoo of QFT complexity metrics, and study whether (or under what circumstances) they satisfy the criteria for being valid complexity metrics. In Section \ref{discri} we discriminate between complexity metrics from different perspectives, the most important one being their hierarchy as lower bounds. We also present conclusive arguments against the use of inhomogeneous costs there. In Section \ref{geomA} we focus on the cost function $F_{\braket{H^2}}$, discussing its relation to the standard Hermitian metric in Hilbert space. We explain how the symplectic structure associated to this metric connects it to coadjoint orbit actions as well as to the actual quantum dynamics of the system. In Section \ref{chaoss}, we describe how the present formulation allows to clarify the relation between complexity and chaos. We close in Section \ref{discuss}. In Appendix \ref{ap1} we present a collection of situations in which the instantaneous Hamiltonian characterizing motion along general unitary trajectories can be computed explicitly. Appendix \ref{ap2} is a more or less self contained account of the theory of coadjoint orbit actions.

\section{A mathematical prelude: the Banach-Lie group of unitaries}
\label{sec:BLgroups}
The set of all unitary transformations of a given quantum system, which we denote $\cM$ from now on, plays a crucial role in the complexity discussion. The goal of this  section is to take a closer look into the mathematical structure of $\cM$ when it is infinite-dimensional  (the finite-dimensional case is classical in the standard theory of Lie groups).  The section is more or less self contained and written in an informal ---yet rigorous--- mathematical language. Less mathematically oriented readers may go straight to Section \ref{bounds}.

Ideally, we would like to endow $\cM$ with the structure of a ``smooth infinite dimensional Lie group'', in order to  be able to do Riemannian geometry on $\cM$ and define the notions of smooth Riemannian metric, Levi-Civita connection and piece-wise smooth geodesic. However, introducing such smooth Lie structure on $\cM$ may or may not be possible depending on the precise meaning we want to grant to the terms ``smooth'' and  ``Lie structure'' in an infinite-dimensional context. Since this issue is generally overlooked in the physics literature, we will discuss it here in some detail by endowing $\cM$ with successive layers of topological and geometric structure. The goal will be to justify, at least formally, the various mathematical manipulations we will perform throughout the rest of the manuscript.

Let $\cH$ be the Hilbert space associated to the quantum system under consideration. Its group of unitary operators is set-theoretically defined as the set of all unitary transformations of the quantum system, with group operation given by composition. As observed in \cite{Kibble:1978tm}, many of the observables of interest in quantum mechanics, notably the Hamiltonian itself, correspond to unbounded operators, and  must therefore be defined on dense subspaces of physical states of $\cH$. Due to this fact, \cite{Kibble:1978tm} considers a fixed dense subspace $\mathcal{D}\subset \cH$ which is assumed to be common to all the observables of the system. With these provisos in mind, the set of all unitary transformations of the quantum system is identified with the set of all unitary operators of $\mathcal{D}$ with product rule given by composition. A moment of thought reveals that every unitary transformation of $\mathcal{D}$ extends to a unique unitary operator on $\cH$ and vice-versa, every unitary operator on $\cH$ restricts to a unitary transformation on $\mathcal{D}$. Hence, we can equivalently define the unitary group of a quantum system simply as the group of all linear and continuous unitary transformations of $\cH$. This defines $\cM$ as an abstract group, which we need to endow now with a topology. There is a large body of literature treating the different topologies that can be defined on $\cM$, exploring in detail their advantages and drawbacks ---see \eg \cite{ReedSimon,Takesaki}. Here we will focus on the two main natural choices: the norm topology and the strong topology.\footnote{The compact-open topology and the weak operator topology on $\cM$ are equivalent to the strong topology and hence we do not need to consider them separately.} We discuss the norm topology  first.

In order to define the norm topology, the first piece of data we need to introduce is the vector space $\cL(\cH)$ of all linear endomorphisms of $\cH$, that is, the vector space of all linear maps from $\cH$ to itself. Denote by $\mathcal{B}(\cH)\subset \cL(\cH)$ the vector space of all bounded linear operators
\begin{equation}
\mathcal{B}(\cH) \equiv \left\{ T\in \cL(\cH): \exists\,\, c>0\,\, |\,\, \forall \,\, \psi \in \cH \,\, \vert T(\psi)\vert_{\cH} \leq c |\psi|_{\cH}  \right\}\, ,
\end{equation}
where $|\cdot |_{\cH}$ is the norm of $\cH$. The ``unitary group'' $\cM$ as defined above is then realized as the subgroup of the unit group of $\mathcal{B}(\cH)$ which preserves the Hilbert structure of $\cH$. Equip $\mathcal{B}(\cH)$ now with the ``operator norm'' $\vert\cdot\vert_{\mathcal{B}}$, which is defined as follows (see also \req{norm} below)
\begin{equation}
\vert T\vert_{\mathcal{B}} \equiv \mathrm{sup}_{\psi\in \mathbb{S}^{\infty}\subset\cH}\vert T(\psi)\vert_{\cH}\, .
\end{equation}
Endowed with this norm, $\mathcal{B}(\cH)$ becomes a Banach space, of which the subspace $\mathfrak{u}(\cH) \subset \mathcal{B}(\cH)$ of skew-Hermitian linear and bounded operators is a Banach subspace when endowed with the induced norm. Furthermore, equipped with the standard Lie bracket defined in terms of composition, $\mathfrak{u}(\cH)$ becomes a Banach-Lie algebra, which is a Banach space equipped with a bilinear and smooth skew-symmetric bracket satisfying the Jacobi identity. The norm topology on $\cM\subset \mathcal{B}(\cH)$ is defined as the subspace topology with respect to the topology on $\mathcal{B}(\cH)$ induced by $\vert \cdot\vert_{\mathcal{B}}$. Equipped with this topology, $\cM$ is a contractible and metrizable topological group which we will momentarily denote by $\cM_n$. Furthermore, it can be shown that the exponential map
\begin{equation}
\mathrm{Exp}\colon \mathfrak{u}(\cH) \to \cM\, , \quad T \to \sum_{n=0}^{\infty} \frac{T^n}{n!} \, ,
\end{equation}
is a local homeomorphism when appropriately restricted to an open neighborhood of zero in the Banach space $\mathfrak{u}(\cH)$ and, in addition, it can be used to endow $\cM$ with the structure of a Banach-Lie group. That is, $\cM$ is a Banach manifold locally modeled on $\mathfrak{u}(\cH)$, with local charts given by the exponential, and such that the composition of unitary operators is smooth. Not only that, $\cM$ becomes a parallelizable, contractible, smooth Banach-Lie group. The Banach-Lie group structure is one of the nicest structures we can expect to construct on an infinite-dimensional group, and allows to intuitively transport to $\cM$ many results and manipulations which hold for finite-dimensional Lie groups. In particular, we can unambiguously talk about smooth metrics, connections, smooth curves and geodesics on $\cM$. Endowing $\cM$ with this smooth Lie-group structure can be used to justify most of the formal manipulations occurring in the rest of the paper. Despite these convenient features, the norm topology has not been favored in the mathematical quantum physics literature, the main reason being its pathological behavior with respect to representations of Lie groups. More precisely, suppose that
\begin{equation}
\Phi\colon G \times \cH \to \cH\, , \quad (g,\psi) \mapsto \Phi_g(\psi)\, ,
\end{equation}
is a continuous unitary representation of a finite-dimensional Lie group $G$. Then, if $\cM$ is equipped with the norm topology, the associated homomorphism $G\to \cM_n$ given by $g\mapsto \Phi_g$ may not be continuous even if the representation is (it is continuous if and only if $G$ is discrete). Furthermore, by Stone's theorem if $\left\{ U_t \right\}_{t\in \mathbb{R}}$ is a group-family of unitary operators (strongly) continuous in $\cM_n$, then it is generated by a bounded skew-Hermitian operator, that is,
\begin{equation}
U_t = \mathrm{Exp}(t T)\, , 
\end{equation} 
with $T\in \mathfrak{u}(\cH)$ bounded.  This rules out the possibility of having continuous families of unitary operators generated by unbounded operators such as the position operator, which is clearly inconvenient for quantum mechanics applications. In conclusion, the norm topology is too fine to do representation theory. These and other reasons traditionally favored the use of the strong topology on $\cM$ instead.

One of the simplest ways to define the strong topology is by specifying which sequences are convergent in this topology. From this point of view, a sequence $\left\{T_k\right\}$ on $\mathcal{B}(\cH)$ is convergent to $T\in \mathcal{B}(\cH)$ in the strong topology if and only if
\begin{equation}
\lim_{k\to\infty} T_k(\psi) = T(\psi)\, , \quad \forall\,\, \psi\in \cH\, .
\end{equation}
The strong topology on $\cM$ is the subspace topology induced by $\mathcal{B}(\cH)$ equipped with the strong topology as defined above. We will momentarily denote by $\cM_s$ the unitary group $\cM$ equipped with the strong topology. Equipped with this topology, $\cM_s$ can be shown to be a metrizable, contractible ---if $\cH$ is infinite dimensional--- topological group for which continuous unitary representations correspond with continuous homeomorphisms $G\to \cM_s$. In particular, continuous families of unitary operators in $\cM_s$ can be generated with unbounded operators. However, the exponential map in this case, although continuous, fails to be a local homeomorphism. Moreover it can be shown that $\cM_s$ cannot be a Banach-Lie group modeled on $\mathfrak{u}(\cH)$. The best we can do is to endow $\cM_s$ with the structure of a Fr\'echet-Lie group. That is, with this structure $\cM_s$ would be locally modeled on open sets on a Fr\'echet space \cite{Neeb}, which is a particular type of topological vector space generalizing the notion of Banach space (a Fr\'echet space is not necessarily equipped with a norm). However,  in contrast to Banach spaces, the standard notion of differentiability occurring in $\mathbb{R}^n$ does not extend naturally to Fr\'echet spaces. Still, a notion of ``directional derivative'', called in the literature ``Gateaux derivative'' \cite{Neeb}, can be defined, and we believe this is enough for some of the manipulations that we shall need later on, such as taking the velocity of a curve in $\cM_s$, to hold. However, it is unclear to us whether the framework of Fr\'echet-Lie groups allows for the type of Riemannian geometry on $\cM_s$ necessary for the continuous approach to complexity we will be dealing with through the rest of the paper.

Summarizing the previous discussion, we conclude the following: i) The norm topology on $\cM$ offers a convenient framework in which $\cM_n$ becomes a contractible and metrizable Banach-Lie group. We can do Riemannian geometry on $\cM_n$ intuitively, as in the finite-dimensional case. However, $\cM_n$ does not behave well with respect to continuous unitary representations and in particular, due to Stone's theorem, with respect to one-parameter families of unitary maps associated to unbounded generators. ii) The strong topology on $\cM$ offers a convenient framework for unitary representations, in which every continuous unitary representation of a finite dimensional Lie group corresponds to a continuous map $G\to \cM_s$. However, $\cM_s$ is not a Banach-Lie group modeled on skew-adjoint endomorphisms of $\cH$ and, in particular, the exponential map is not a local homeomorphism. It is not clear if the (contractible and metrizable) Fr\'echet-Lie group structure existing on $\cM_s$ is enough for the requirements of the complexity theory to be developed in later sections. It is important to remark that (strongly) continuous one-parameter families of unitary operators $U(s)$ in the norm topology are always given by the exponential of bounded operators. Hence, if we have a physical argument to consider exclusively bounded observables, then the norm topology is as good as the strong topology with regards to their behavior with respect to continuous families of unitary operators. 
 It is beyond the scope of this paper to determine in a mathematically rigorous way which of the topologies is the appropriate one  for the continuous complexity approach (if any). We leave the question open for debate for the moment.

\section{Geometric lower bounds to computational complexity}\label{bounds}
As anticipated in the previous section, our starting point in the complexity discussion is the set of all unitary transformations $\cM$ of a given quantum system, which we will consider, following that section, either as a smooth Banach-Lie group or as a Fr\'echet-Lie group depending on which shall be the most convenient choice. No further mention will be made to this mathematical point and we will assume that all computations hold either in the Banach-Lie group framework or in the Fr\'echet-Lie group one.

 If any unitary $U_f \in \mathcal{M}$, can be reached from the identity operator $\mathds{1}$ by successive application of a certain number of operators of a given set $\mathcal{G}\subset \mathcal{M}$, such set is called ``universal'' in the complexity context. The elements of $\mathcal{G}$ are called ``gates'' and a product of gates giving rise to $U_f$ is called a ``protocol''. Given a universal set of gates $\mathcal{G}$, the smallest number of elements of such a set required to achieve $U_f$ from $\mathds{1}$ is called the ``complexity'' of the computation, which we denote $\mathcal{C}_{\mathcal{G}}(U_f)$.

In general, there exist infinitely many different protocols producing $\mathds{1}\overset{\mathcal{G}}{\rightarrow} U_f$. A canonical notion of ``computational cost'' can be associated to each of them. The cost of a given protocol is nothing but the total number of gates it involves  (repeated gates add one to the cost every time they appear). Then, complexity can be thought of as the minimal total cost required for achieving the computation  $\mathds{1}\overset{\mathcal{G}}{\rightarrow} U_f$. 

The question arises on how to find such optimal protocols or, at least, on how to get good approximations to their associated complexity. An interesting approach to this problem was put forward by Nielsen \emph{et al.} \cite{2005quant.ph..2070N,2006Sci...311.1133N,2007quant.ph..1004D}. The idea is to switch from discrete to continuous protocols  and associate computational costs to the latter in a way such that their corresponding optima produce lower total costs than their discrete counterparts ---\ie they provide lower bounds to the actual complexity of the protocol. Continuous computational costs or ``complexity metrics'' satisfying this requirement ---which depends on the choice of gate set $\mathcal{G}$--- are said to be ``$\mathcal{G}$-bounding''.\footnote{In the holographic complexity literature, the subtle dependence of the complexity metric on the gate set $\mathcal{G}$ has not been considered properly.} This notion was originally conceived for spin systems \cite{2005quant.ph..2070N} but, as we show here, it can be naturally extended to generic quantum systems, including QFTs.

We start this section reviewing the role played by the so-called ``instantaneous Hamiltonian'', which defines infinitesimal motion along a given curve on the unitaries manifold. Then we turn to the problem of defining appropriate notions of  cost measures, and we review the different proposals presented in the literature so far, and the relations between them.
While discussing the differences between the different measures will be the subject of the following sections, we do comment here on how they deal with a certain gauge ambiguity present in the definition of instantaneous Hamiltonians. Then, we explain the $\mathcal{G}$-bounding condition and how and under what conditions cost measures satisfying it provide lower bounds to computational complexity.

\subsection{Continuous protocols and instantaneous Hamiltonians}\label{contpro}
Pairs of elements of $\mathcal{M}$, $\{ U_0,U_f\}$ can be connected by infinitely many continuous paths. Any of such paths can be parametrized by some affine parameter $s$ so that $U(s)$ is the intermediate unitary along the curve corresponding to that value of the affine parameter, and $U(0)=U_0$, $U(1)=U_f$. From now on, we will  always be considering paths in  $\mathcal{M}$ which start at the identity operator, namely, $U(0)=U_0=\mathds{1}$. Given some curve $U(s)$ in $\cM$, it is useful to introduce the ``instantaneous Hamiltonian'' as the Hermitian operator $H(s)$ such that 
\begin{equation}\label{u0}
U(s+ ds)=e^{-iH(s) ds } U(s)\, ,
\end{equation}
namely, the one which generates infinitesimal motion along the curve parametrized by $s$. From a more geometric point of view, we can identify $-i H(s)$ with the Maurer-Cartan form of $\cM$ evaluated at the velocity of $U(s)$. Given a curve $U(s)$ in $\cM$ denote by:
\begin{equation}
\dot{U}(s_0) = \lim_{\varepsilon\to 0} \frac{U(s_0 + \varepsilon) - U(s_0)}{\varepsilon}\, ,
\end{equation}
the derivative of $U(s)$ at $s_0$, whenever it exists. Assuming it exists, the previous equation is equivalent to:
\begin{equation}
\label{instMC}
-i H(s_0) = \dot{U}(s_0) U(s_0)^{-1}\, ,
\end{equation}
upon using Equation \eqref{u0}. The right hand side is nothing but the Maurer-Cartan form evaluated at $\dot{U}(s_0)$ (and defined with respect to the right action of $\cM$ on itself). Instantaneous Hamiltonians $H(s)$ play a protagonist role in essentially any possible geometric approach to complexity ---one which has not been acknowledged very often in the literature.

Given some reference state $\ket{\psi(0)}$, we can define $\ket{\psi(s)}$ as the one resulting from acting with $U(s)$ on it, namely $\ket{\psi(s)}\equiv U(s)\ket{\psi(0)}$. Then, $H(s)$ moves us from $\ket{\psi(s)}$ to $\ket{\psi(s+ds)}$ through \req{u0}. Expanding in both sides of \req{u0}, it follows that 
\begin{equation}\label{sch}
i\dot U(s)= H(s) U(s) \quad \Leftrightarrow \quad i \frac{d}{ds}\ket{\psi(s)}= H(s)\ket{\psi(s)}\, ,
\end{equation}
which is nothing but Schr\"odinger's equation and, in principle, allows us to construct $H(s)$ once we know the curve $U(s)$. Alternatively, we can express $U(s)$ as a function of $H(s)$ as the path-ordered integral
\begin{equation}\label{usp}
U(s)=\mathcal{P} e^{-i \int_{0}^s H(s') ds'}\, .
\end{equation}
These relations may look innocent at first sight, but they are not in general. Imagine we can construct $U(s)$ from some Hermitian operator $\mathcal{O}(s)$ as\footnote{It is illustrative to compare the action of $e^{-i\mathcal{O}(s)}$  on the initial state  $\ket{\psi(0)}$ with the one of the instantaneous Hamiltonian. While  $e^{-i\mathcal{O}(s)}\ket{\psi(0)}=\ket{\psi(s)}$ for any finite $s$, we can only say that the action of $e^{-iH(0) ds}$, on the same state moves us to the infinitesimally nearby state  $\ket{\psi(ds)}$.} 
\begin{equation}\label{OO}
U(s)=e^{-i\mathcal{O}(s)}\, ,
\end{equation}
where one would expect $\mathcal{O}(s)$ to be expandable in some basis of Hermitian operators,  schematically: $\mathcal{O}(s)=\sum_I \theta_I(s)K_I$ and $\mathcal{O}(s)=\int dk \theta_k(s) K(k)$
 in the discrete and continuous cases respectively. Then, using \req{u0} and  \req{OO} we can write 
\begin{equation}\label{usds}
U(s+ ds)=e^{-i \left[ \mathcal{O}(s) +\frac{d \mathcal{O}(s) }{ds}\right]}=e^{-iH(s) ds } e^{-i\mathcal{O}(s)}\, .
\end{equation}
As previously observed \cite{2005quant.ph..2070N,Magan:2018nmu}, the result for $H(s)$ turns out to involve an infinite sum of nested commutators, namely\footnote{The first few terms read 
\begin{equation}
H(s)=\frac{d \mathcal{O}(s) }{ds}-\frac{i}{2!}\left[\mathcal{O}(s) ,\frac{d \mathcal{O}(s) }{ds}\right] -\frac{1}{3!}\left[\mathcal{O}(s),\left[\mathcal{O}(s) ,\frac{d \mathcal{O}(s) }{ds}\right]\right]+\dots
\end{equation}
}
\begin{equation}\label{nested}
H(s)=\sum_{j=0}^{\infty} \frac{(-i)^j}{(j+1)!}\,  {\rm ad}^j_{\mathcal{O}(s)}\left[\frac{d \mathcal{O}(s) }{ds}\right]\, , \quad \text{where}\quad   {\rm ad}^j_{A}\left[B\right]\equiv [ \underbrace{A,[A,[A,[\dots ,[A}_{j}, B]\big]\, .
\end{equation}
This is very difficult to deal with in general.\footnote{As mentioned above, we can expand $\mathcal{O}(s)$ in some basis of Hermitian operators $K_I$ or $K(k)$, and \req{nested} will ultimately give rise to an analogous  sum/integral for the instantaneous Hamiltonian, $H(s)=\sum_I Y^I(s) K_I$ or $H(s)=\int dk Y_k(s) K(k)$. In that case, the challenge would reside in computing the $Y^I(s)$ or the $Y_k(s)$ in the discrete and continuous cases, respectively. Naturally, any physically meaningful notion will better be independent of the basis chosen. We postpone a discussion regarding this issue to the following section.} In certain special situations, however, the sum can be performed explicitly. We review some of this cases in Appendix~\ref{ap1}. Essentially, to the best of our knowledge, all protocols considered so far in the complexity literature fall within one of the classes considered there (although in a sometimes obscurely presented way). A new important example consisting of unitary curves generated by generalized free fields is also included in that appendix.

In sum, given a ``big'' unitary $U(s)$ written in terms of some Hermitian operator $\mathcal{O}(s)$ as in \req{OO}, the instantaneous Hamiltonian $H(s)$ can be obtained  (at least in principle) from the infinite sum of nested commutators in \req{nested}.

Having defined continuous protocols, we need now to associate computational costs to them. This can be achieved by associating a cost density $F(H(s))$ to the instantaneous Hamiltonians $H(s)$ characterizing continuous paths on $\mathcal{M}$. Integrated along a given curve, this cost density would yield the total cost, namely
\begin{equation}\label{totcos}
C_F(U_f)=\int_0^1 ds \, F\left[U(s),H(s) \right]\, , \quad \text{where}\quad U(0)\equiv \mathds{1}\, , \, U(1)\equiv U_f\, .
\end{equation}
Since, as remarked earlier, $H(s)$ can be interpreted as the Maurer-Cartan form evaluated at the tangent space of $U(s)$, we will interpret $F[U(s),H(s)]$ as being the evaluation at $\dot{U}(s)$ of a continuous non-negative function defined on $T\cM$. More precisely, we set:
\begin{equation}
F[U(s),H(s)] \equiv \hat{F}[U(s),\dot{U}(s)]\, ,
\end{equation}
for an appropriately defined continuous function $\hat{F}\colon T\cM \to [0,\infty)$ on $T\cM$ and for all $U(s)$ (for ease of notation we will drop the \emph{hat} on $F$ in the following). The hope is now that by minimizing the cost functional for sensible notions of  continuous non-negative functions $F$, we can estimate the complexity of the computation, in the sense of producing as good as possible lower bounds on $ \mathcal{C}_{\mathcal{G}}(U_f)$ ---see subsection \ref{Gbp}. 

Notice that irrespective of whether $F$ provides actual lower bounds to quantum complexity, there are a number of sensible generic assumptions one can make regarding its properties \cite{2005quant.ph..2070N}: i) continuity; ii) positivity, $F\left[U(s),H(s) \right]\geq 0$ with equality iff $H(s)=0$; iii) positive homogeneity, $F\left[U(s),\alpha H(s) \right]= \alpha F\left[U(s),H(s) \right]\geq 0$ for any $\alpha \in \mathbb{R}^+$ and any $H(s)$; iv) triangle inequality, $F\left[U(s),H_1(s) \right]+ F\left[U(s),H_2(s) \right]\geq F\left[U(s),H_1(s)+H_2(s) \right] $; v) the Hessian of $F$ is positive definite. Assuming these conditions for $F$, the pair $(\cM,F)$ becomes a ``Finsler manifold'' with Finsler structure $F$.


\subsection{Cost measures in quantum field theory}
Many different classes of cost measures $F$ have been considered in the literature so far. They can be classified attending to different criteria: i) whether or not, in addition to the instantaneous Hamiltonian $H(s)$, they depend on the instantaneous state $\ket{\psi(s)}$; ii) whether or not they are invariant under changes in the basis of Hermitian operators in which we expand the instantaneous Hamiltonian; iii) whether they satisfy the positive homogeneity property described above, or rather, they are homogenous of degree $p\neq 1$, namely, $F[U(s),\alpha H(s)]=\alpha^p F[U(s), H(s)] $ for any $\alpha \in  \mathbb{R}^+$. We will refer to the latter as ``inhomogeneous'' costs henceforth. In this subsection we present a summary of the different costs presented so far, including a new one ---see \req{norm} below. 
We are succinct here regarding which cost functions are ``better'' than others. That issue will be the main focus of the remainder of the paper. Nevertheless, we do make some comments regarding basis-dependent costs and also,  in Section \ref{gaugis}, about the different ways in which the various costs deal with the gauge ambiguity inherent to the definition of the instantaneous Hamiltonian. 

Observe that all cost functions presented here can be modified by including weights (penalty functions) which would discriminate certain directions on $\mathcal{M}$. We will discuss the role of those in Section \ref{penaltyyexpulsion}, including our proposal on how to deal with them on general grounds.

\subsubsection{State-independent measures}
Let us consider first the case of cost functions which do not depend on the instantaneous state  $\ket{\psi(s)}$.
As we mentioned earlier, when the dimensionality of the set of relevant Hermitian operators is finite ---either because the system has a finite number of degrees of freedom, or because we are restricting the allowed operations to a finite dimensional submanifold of $\mathcal{M}$--- the instantaneous Hamiltonian can be written in some basis of such operators as $H(s)=\sum_I Y^I(s) K_I$ for certain coefficients  $Y^I(s)$ to be computed from \req{sch} (or equivalently, \req{nested}).
In that context and for spin systems, Nielsen  proposed the cost functions
   \cite{2005quant.ph..2070N}
\begin{equation}\label{Fniel}
F_1\equiv\sum_I |Y^I(s) |\, , \quad \, \, \,\, \,F_2\equiv\sqrt{\sum_I \left(Y^I(s) \right)^2}\, ,
\end{equation}
which are both homogeneous of degree $1$.
 An inhomogeneous proposal was suggested in \cite{Jefferson:2017sdb}
\begin{equation}\label{FRob}
F_{\kappa }\equiv\sum_I |Y^I(s) |^{\kappa}\, .
\end{equation}
The motivation for introducing the family $F_{\kappa}$ with $\kappa>1$ was to produce a qualitative agreement between the orders of the leading UV divergences resulting from the holographic ``complexity equals action'' \cite{Brown:2015lvg,Brown:2015bva} and ``complexity equals volume''  \cite{Susskind:2014rva,Stanford:2014jda} proposals and the one resulting from the continuous limit of a lattice of coupled harmonic oscillators (the setup in \cite{Jefferson:2017sdb} involved Gaussian states as initial and target states and a particular choice for the subset of Hermitian operators allowed to appear in $H(s)$). While the holographic result had a $\sim V/\delta^{(d-1)}$ dependence \cite{Carmi:2016wjl}, the free-scalar result turned out to be $\sim \left(V/\delta^{(d-1)}\right)^{1/2}$. 
In Section \ref{discri} we explain the origin of this, and other previously found apparent mismatches, and why using inhomogeneous costs in order to make both scalings agree is unjustified. Naturally, another class of homogeneous measures constructed out from $F_{\kappa}$ are simply given by $\left[F_{\kappa }\right]^{1/\kappa}$.

There is an obvious drawback inherent to all the above cost functions. This is the fact that they depend on the basis of Hermitian operators in which one expands the instantaneous Hamiltonian. In other words, they give an ambiguous answer for the cost associated to a given instantaneous Hamiltonian: expanding  $H(s)=\sum_I Y^I(s) K_I$ and $H(s)=\sum_I \tilde Y^I(s) \tilde K_I$ in two different basis, we would obtain two different answers for each of the cost functions defined above, which does not seem acceptable. 

In principle, one can declare that the  $Y^I(s)$ appearing in the above cost functions are the ones corresponding to some particular basis selected by certain criterion. A somewhat canonical possibility would be to  choose (whenever possible) a basis of unit-trace Hermitian operators such that 
\begin{equation}\label{Hermis}
 {\rm Tr} \left[K_I K_J\right]=\delta_{IJ}\, ,
 \end{equation}
 up to an overall constant which can be chosen to be the dimensionality of the Hilbert space. 
Written in such a basis, some of the above cost functions equal certain basis-independent costs defined directly in terms of the instantaneous Hamiltonian. For example,
\begin{align}\label{trH2}
F_{{\rm Tr}H^2}\equiv \sqrt{ {\rm Tr} \left[H(s)^2\right]}&=\sqrt{\sum_{I,J} Y^I(s) Y^J(s) {\rm Tr} \left[K_I K_J\right]}=F_2\, .
\end{align}
Namely, the basis-independent cost function $F_{{\rm Tr}H^2}$ defined above agrees with $F_2$ when the latter is understood as being referred to a basis of  Hermitian operators satisfying \req{Hermis}. This is for example what Nielsen does in \cite{2005quant.ph..2070N} for $n$-qubit systems ---and it seems to be a more or less standard choice in that context. In particular, one chooses a basis of Hermitian operators consisting of $n$-fold tensor products of single-qubit Pauli matrices and the identity operator, with the exception of\footnote{For instance, for $n=3$ examples of basis elements would be $ \sigma_x \otimes \mathds{1} \otimes \sigma_z$ or $\mathds{1} \otimes \mathds{1}  \otimes \sigma_y$.}  $K=\mathds{1}^{\otimes n} $ ---see Section \ref{gaugis} below. With this definition, there are $4^n-1$ generators, and they precisely satisfy \req{Hermis} up to an overall factor. On general grounds, with the exception of those situations in which the chosen basis is such that the resulting cost function is in fact basis-independent ---like in the case just described--- notions of complexity relying on basis-dependent measures are ill-defined and henceforth should not be considered.

Naturally, one can propose other basis-independent cost functions such as
\begin{equation}\label{trH}
F_{|{\rm Tr}H|}\equiv  | {\rm Tr} H | \, , \quad F_{\rm Sch}\equiv \left[{\rm Tr} \left((H^2)^{\frac{p}{2}}\right)\right]^{\frac{1}{p}}\, ,
\end{equation}
the second of which is usually known as the Schatten norm and, in the context of QFT complexity, it was precisely introduced in \cite{Hackl:2018ptj,Guo:2018kzl} as an alternative basis-independent notion to $F_{1}$, $F_2$ and $F_{\kappa}$.

A new state-independent cost can be constructed by considering the norm of the instantaneous Hamiltonian, namely 
\begin{equation}\label{norm}
F_{||H ||} \equiv ||H(s) || \, , \quad \text{where} \quad ||H(s) || \equiv \underset{\ket{\psi} \in \mathcal{H}}{{\rm sup} }\sqrt{ \frac{\braket{\psi | H(s)^2 |\psi}}{\braket{\psi|\psi}}} \, .
\end{equation}
The cost associated to the instantaneous Hamiltonian would be given by its norm. We will see later that $F_{||H ||}$ has some interesting properties. Before doing so, let us continue with our catalog of cost functions. We turn now to state-dependent measures.

\subsubsection{State-dependent measures}
\label{statedep}
Observe that cost functions involving traces of powers of the instantaneous Hamiltonian can be thought of as corresponding to expectation values of such operators on the totally mixed state $\rho_{\rm mixed}=\mathds{1}/{{\rm dim} \mathcal{H}}$ \cite{Magan:2018nmu}. Both from a physical and an operational point of view, it is only natural to replace $\rho_{\rm mixed}$ by the actual instantaneous state of the system $\ket{\psi(s)}=U(s) \ket{\psi (0)}$, which leads to the notion of state-dependent costs. In addition, from a QFT perspective, state-dependent costs have the advantage of being finite, while traces in QFT are generically divergent,\footnote{Actually, in local and continuum QFT, as described by its algebraic formulation, the algebras involved are of type-III, and traces do not even exist for them. This is because such algebras only contain projectors of zero or infinite dimensionality \cite{Haag:1992hx}.} and therefore force us to choose directly bounded operators.\footnote{Here we would like to emphasize that in spite of the typically divergent cost associated to generic protocols when using $\rho_{\rm mixed}$ in QFT, we expect that the fact that we will be minimizing over the possible unitary paths will select contributions to the instantaneous Hamiltonian corresponding to bounded operators with finite traces. We say more about this below.}

To the best of our knowledge, the first example of state-dependent cost considered in the complexity literature was the Fubini-Study metric \cite{Chapman:2017rqy}. This is defined in terms of the instantaneous Hamiltonian as
\begin{equation}\label{FS}
F_{\rm FS}\equiv \sqrt{ \braket{\psi(s)|  H(s)^2 |\psi(s)} - \braket{\psi(s)| H(s) |\psi(s)}^2} \, ,
\end{equation}
which is nothing but the square root of its variance, $F_{\rm FS}=\sigma_{\ket{\psi(s)}}( H(s))$. Two other measures related to the former have also been proposed \cite{Magan:2018nmu},
\begin{equation}\label{H2Ja}
F_{\braket{H^2}}\equiv \sqrt{ \braket{\psi(s)|  H(s)^2 |\psi(s)} }\, , \quad F_{|\braket{H}|}\equiv | \braket{\psi(s)|  H(s) |\psi(s)} |\, .
\end{equation}
Observe that these two can be obtained from $F_{{\rm Tr}H^2}$ and $F_{|{\rm Tr} H|}$  by replacing $\rho_{\rm mixed}$ by the instantaneous state of the system, as anticipated above. Note also that $F_{\rm FS}^2=F_{\braket{H^2}}^2-F_{|\braket{H}|}^2$. 
In Section \ref{geomacQM}, we will comment on the relevant geometric status of $F_{\rm FS}$ and specially $F_{\braket{H^2}}$.

The next member of this cost-measure catalog is given directly by the variance of the instantaneous Hamiltonian, namely 
\begin{equation}
F_{\sigma^2}\equiv \braket{\psi(s)|  H(s)^2 |\psi(s)} - \braket{\psi(s)| H(s) |\psi(s)}^2=F_{\rm FS}^2\, .
\end{equation}
This inhomogeneous cost function was suggested in \cite{Belin:2018bpg}. There, $F_{\sigma^2}$ was presented pulled back to a basis of holomorphic coherent states of the Hilbert space, but it can be of course considered more generally. In \cite{Belin:2018bpg}, $F_{\sigma^2}$ was proposed in order to reproduce the volume of extremal slices in the bulk in some holographic examples. Just like for the $F_{\kappa}$ introduced above, we comment on the unsuitability of $F_{\sigma^2}$ as a possible notion of complexity and on the origin of the square-root scaling (apparently requiring the use of $F_{\sigma^2}$ instead of $F_{\rm FS}$ or $F_{\braket{H^2}}$) in Section \ref{discri}.\footnote{In favor of these measures and in other situations, it has been argued that complexity measures need to be ``additive''. This is a confusion which is in fact explained in the original complexity geometry paper \cite{2005quant.ph..2070N}. If we have two systems, and gates are not allowed to contain operators that are products of one system with the other, then all measures we have been discussing are additive. If, on the other hand, we allow such gates, there is no reason to expect additivity, because when adding such gates we are opening ways for computation that might decrease the additive result, and that were not allowed before. Also, from a physical perspective, interacting systems are those for which actions or energies are not additive.}

Yet another state-dependent measure was recently presented in \cite{Yang:2019udi,Yang:2019iav}. The authors of those papers suggest using a notion of complexity given by
\begin{equation}\label{yang}
C=-2\log |\braket{\psi _f |\psi_0}|\, ,
\end{equation}
where $\ket{\psi_f}= U_f \ket{\psi_0}$ is the final state. This notion only depends on the initial and final states, but not on the path, so it does not correspond to a functional of the form \req{totcos} for any local metric $F$. While, by construction, \req{yang} typically becomes an on-shell action in the semiclassical limit \cite{Yang:2019udi}, which looks appealing, it does not seem to provide a reasonable notion of complexity. This can be seen, \eg by observing that  \req{yang} associates an infinite cost  to any protocol connecting any pair of orthogonal states. This is indeed discussed in \cite{Yang:2019udi}. We will see later on in Section \ref{geomA} that $F_{\braket{H^2}}$ can also provide a realization of the ``complexity equals action'' idea while providing a reasonable lower bound for complexity.

Note that if we were to consider \req{yang} only infinitesimally ---\ie as a notion of distance between nearby states--- then it would be related to $F_{\sigma^2}$ as defined above. In order to see this, let $\ket{\psi_f}=\ket{\psi(s+ds)}$ and $\ket{\psi_0}=\ket{\psi(s)}$, which are therefore taken to be infinitesimally close to each other. Then, for some $H(s)$, $\ket{\psi_f}=e^{-i H(s) ds}\ket{\psi_0}=\left[\mathds{1}-iH(s)ds-\frac{1}{2}H(s)^2 ds^2+\dots \right]\ket{\psi_0} $. It follows that
\begin{equation}
|\braket{\psi_f |\psi_0}|=\sqrt{1-F_{\sigma^2} ds^2} \quad \Rightarrow \quad -2\log |\braket{\psi _f |\psi_0}|= F_{\sigma^2} ds^2\, ,
\end{equation}
which sort of gives an infinitesimal cost, but only sort of, because the arc length appears squared and the corresponding quantity cannot be properly integrated unless one takes the square-root of the whole expression. Note that the Fubini-Study cost function can be similarly obtained from $C={\rm arccos} |\braket{\psi_f |\psi_0}|$, which infinitesimally gives $C=F_{\rm FS} ds$ ---and can be properly integrated as it is.


\subsubsection{Gauge ambiguities in the instantaneous gate}
\label{gaugis}

Given a curve $U(s)$ in the unitary manifold $\cM$, the associated instantaneous Hamiltonian is uniquely defined by equation \eqref{instMC}. On the other hand, given a fixed curve $U(s)$ and an initial state $\ket{\psi_{0}} \in \cH$, we obtain a curve $\ket{\psi(s)}$ on $\cH$ defined as $\ket{\psi(s)} = U(s)\ket{\psi_0}$ and whose evolution is unitary. It is sometimes convenient to work with curves in $\cH$ rather than with curves in $\cM$. Notice that, while $U(s)$ determines $\ket{\psi(s)}$, the opposite will not be true in general due to the fact that $\ket{\psi(s)}$ will have non-trivial stabilizer for all values of the protocol time. 

In this context, it is natural to attempt to construct $H(s)$ directly from a curve $\ket{\psi(s)}$ with unitary evolution. However, starting exclusively with $\ket{\psi(s)}$, its associated instantaneous Hamiltonian is ambiguously defined. One way to see this is noting that physical states in quantum mechanics are rays in a Hilbert space: no experiment can distinguish $\vert \psi (s)\rangle$ from $\vert \tilde{\psi} (s)\rangle=e^{i\phi (s)}\vert \psi (s)\rangle$ so both of them represent the same state. Each of the previous vectors has its own instantaneous gate, both of them related through
\begin{equation}\label{gau}
\tilde{H}(s)=H(s)-\dot{\phi}(s)\, .
\end{equation}
This can be easily seen by noting that
\begin{equation}
i \frac{d}{ds}\ket{\tilde \psi(s)}=i e^{i\phi(s)}\left[ i\dot \phi(s)+\frac{d}{ds}\right] \ket{\psi(s)} =  \left[H(s)-\dot\phi(s)\right]\ket{\tilde \psi(s)}\, .
\end{equation}
Eq. (\ref{gau}) is a $U(1)$ gauge symmetry, being $H(s)$ its one-form gauge potential. Since computational costs are functionals of $H(s)$, one could worry about an important loophole in the discussion. Luckily, this is not the case for any metric, albeit for different reasons, which we discuss now.

The clearest case is the Fubini-Study metric (the same applies to $F_{\sigma^2}$), which is exactly constructed so as to avoid such ambiguities. Indeed, using \req{FS} one finds
\begin{equation}
\tilde F_{\rm FS}^2=\braket{ \left[H(s)-\dot\phi(s)\right]^2}-\braket{ \left[H(s)-\dot\phi(s)\right]}^2=F_{\rm FS}^2\, ,
\end{equation}
where $\tilde F_{\rm FS}$ means evaluation in the state $\ket{\tilde{\psi} (s)}$.
This makes the Fubini-Study a natural metric in projective space.  

As it turns out, the other metrics are also insensitive to this ambiguity, since they all give zero to the identity gate by definition. In other words, the cost of doing nothing is zero. Practically, whenever we have an instantaneous gate $H(s)$, we first need to extract from it the term proportional to the identity, and then apply the appropriate cost measure. This procedure ---which, \eg in the case of $n$-qubit systems means removing the contribution proportional to $K=\mathds{1}^{\otimes n} $--- provides an answer which is obviously invariant under gauge redefinitions.

It is important to remark that the way $F_{\rm FS}$ and $F_{\sigma^2}$ deal with the ambiguity is 
 markedly different from the rest. This is beacuse in forcing a zero answer to such gauge transformations, they also give zero cost to non-zero operators in certain cases. For example, if we apply $e^{i\sigma_{z}}$ to $\vert\uparrow\rangle$, $F_{\rm FS}$ and $F_{\sigma^2}$ would give zero cost, even though we are actually doing some effort in applying a transformation which is definitely not the identity. More generally, given a state at a given time, $F_{\rm FS}$ and $F_{\sigma^2}$ not only give zero cost to the identity transformation, but to the full stabilizer group of the corresponding state. This implies that such metrics do not allow us to distinguish between circuits doing different numbers of mistakes. In particular, we can construct increasingly long circuits, with any wanted number of gates different from the identity, and still with fixed $F_{\rm FS}$ and $F_{\sigma^2}$ distances. This obviously includes the cases of minimal geodesics. Therefore, minimizing the $F_{\rm FS}$ and $F_{\sigma^2}$ metrics is a strongly degenerate problem from the point of view of the unitary manifold $\mathcal{M}$, since there are infinitely many protocols driving us through the same set of states. We will see later that this is related to the fact that  $F_{\rm FS}$ generically gives rise to worse lower bounds on complexity than other choices such as $F_{\braket{H^2}}$. 

The comments in the previous paragraph do not apply to the rest of state-dependent metrics reviewed before. Those can distinguish the identity operator from elements of the stabilizer group of the state being transformed. For instance, a unitary transformation of the form $e^{-i H}$ acting on an instantaneous state $\ket{E}$ which is an eigenstate of $H$ with eigenvalue $E$ would have costs $F_{\braket{H^2}}=F_{|\braket{H}|}=E$. The minimization of some of those metrics provides unique optimal protocols and, generically, the computational costs assigned to each protocol is bigger than their Fubini-Study counterparts, providing better lower bounds for complexity. We will see this in various explicit situations below. Naturally, the (basis-independent) state-independent metrics $F_{{\rm Tr}H^2}$, $F_{|{\rm Tr}H|}$, $F_{\rm Sch}$, $F_{||H ||}$ ---which by definition do not depend on the instantaneous state--- can naturally distinguish between elements of the stabilizer group as well. For those, no particular simplification occurs for the cost associated to unitaries belonging to the stabilizer group of the instantaneous state.  

Note the previous discussion implicitly assumes that the norms discussed (more generally positive continuous functions) are defined on the tangent bundle $T\cH$ of the Hilbert space $\cH$. However, in Section \ref{contpro} we introduced norms in the context of the manifold of unitary transformations $\cM$. This is not problematic when it comes to computing the associated cost functional. This is because, given a curve $\ket{\psi(s)}$ with unitary evolution, we can in principle always compute its associated class of unitary curves $U(s)$ and then it is to be expected that the given norm $\hat{F}$ on $\cM$ does not depend on the representative chosen, as it happens in the examples above. 

Let us now see what all these cost functions have to do with complexity, understood as the minimum number of gates, belonging to some universal set, required for implementing a given unitary operation. That is the goal of the following subsection.

\subsection{$\mathcal{G}$-bounding condition}\label{Gbp}
Given a set of universal gates $\mathcal{G}$, the total cost of a given computation, $d_F(U_f)$, defined as the minimum of $C_F(U_f)$ ---see \req{totcos}--- for all possible unitary paths in \req{totcos}, can be used to approximate the actual complexity of the computation whenever the corresponding cost function $F$ satisfies the so-called ``$\mathcal{G}$-bounding property''.  A cost function $F$ is $\mathcal{G}$-bounding if it satisfies  
\begin{equation}\label{gbounding}
F\left[U,h \right]\leq 1\, , \quad  \text{for all} \quad e^{-i h} \in \mathcal{G}\, ,
\end{equation}
and for any $U \in \mathcal{M}$. In words, the $\mathcal{G}$-bounding property requires the cost density associated to every universal gate belonging to  $\mathcal{G}$ to be bounded by $1$. This notion was introduced by Nielsen in \cite{2005quant.ph..2070N} in the context of $n$-qubit circuits, for which $\mathcal{M}=SU(2^n)$, and we generalize it here to arbitrary quantum systems.

It follows that for any given $\mathcal{G}$-bounding cost function,
\begin{equation}\label{dfC}
d_F(U_f)\leq \mathcal{C}_{\mathcal{G}}(U_f)\, , \quad \text{for any} \quad U_f \in \mathcal{M}\, .
\end{equation}
Namely, the distance between $\mathds{1}$ and $U_f$ associated to the notion of distance defined by $F$ is a lower bound for the actual number of gates required to produce the computation. 

In order to prove \req{dfC}, we can start with the setup considered in  Section \ref{piece}. Namely, we assume that $U_f$ can be optimally constructed  as in \req{usis} from a finite sequence of unitaries belonging to some universal set, $e^{-i h_{j}}\in \mathcal{G}$. If the optimal protocol requires $N_{\mathcal{G}}$ gates, the complexity of the computation is simply $\mathcal{C}_{\mathcal{G}}(U_f)=N_{\mathcal{G}}$. As we explain in Section \ref{piece}, we can define a curve $U(s)$ connecting the gates involved through \req{usis2}. The resulting curve, whose instantaneous Hamiltonian is given by $H(s)=N_{\mathcal{G}} h_j$ for values of the affine parameter $(j-1)/N_{\mathcal{G}}<s<j/N_{\mathcal{G}}$ is not smooth, as it jumps when $s N_{\mathcal{G}}$ takes integer values. The idea is to regularize this curve by introducing a real-valued smooth function $r(s)$ such that \cite{2005quant.ph..2070N}: i) $r(s)=0$ for $s N_{\mathcal{G}}\in \mathbb{N}$; ii) $r(s)\geq 0$; iii) for any $j\in \mathbb{N}$, 
\begin{equation}\label{rss}
\int_{j/N_{\mathcal{G}}}^{(j+1)/N_{\mathcal{G}}}  r(s)\, ds= \frac{1}{N_{\mathcal{G}}} \, .
\end{equation}
Then, we define a modified instantaneous Hamiltonian as $H_1(s)\equiv r(s) H(s)$ which now corresponds to a different curve $V(s)$ connecting $\mathds{1}$ with $U_f$ ---for a given function $r(s)$, this can be obtained solving \req{sch}. As opposed to the one associated to $H(s)$, the new curve is smooth, and its associated cost is given by 
\begin{align}\label{proof}
C_F=\int_0^1 ds F\left[V(s), H_1(s) \right]=\int_0^1 ds r(s) F\left[V(s), H(s) \right]\leq \int_0^1 ds r(s) N_{\mathcal{G}} = N_{\mathcal{G}}\, .
\end{align}
The second equality makes use of the positive homogeneity of $F$ described above; the inequality uses the fact that $e^{-iH(s)/N_{\mathcal{G}}}\in \mathcal{G}$, the $\mathcal{G}$-bounding property and again the positive homogeneity of $F$; and the last equality uses \req{rss}. Since $d_F(U_f)$ is the minimum of all possible $C_F(U_f)$, it is, in particular, smaller than the cost assigned to the curve $V(t)$ defined above, and hence \req{dfC} holds.

This proof is very similar to the one presented by Nielsen in \cite{2005quant.ph..2070N} for $\mathcal{M}=SU(2^n)$, but note that it holds on general grounds as long as the hypothesis introduced above are satisfied.
The key aspect is that, given a set of universal gates $\mathcal{G}$, the candidate cost function satisfies the $\mathcal{G}$-bounding property \req{gbounding}.

Let us now analyze under what conditions the cost functions introduced above provide actual lower bounds for the complexity of a given computation. In order for this to be the case, given a universal set of gates $e^{-ih_{(j)}}\in \mathcal{G}$, each cost is required to satisfy the $\mathcal{G}$-bounding property in \req{gbounding}. 
There seem to be two different approaches. 
\begin{enumerate}
\item Given a fixed cost function $F$, find the possible sets of universal gates $\mathcal{G}$ such that $F$ satisfies the $\mathcal{G}$-bounding property \req{gbounding}. $F$ will  provide a lower bound for complexity only with respect to those sets satisfying \req{gbounding}. If we adopt this perspective, the $\mathcal{G}$-bounding condition should not be understood as a condition on the cost function, but  as a constraint on the possible families of universal gates with respect to which a given cost defines a lower bound of complexity. In the case of homogeneous state-dependent costs, a reasonable criterion would be to consider gates such that $||h_{(j)} ||\leq 1$ ---as defined in \req{norm}--- for all $j$. Then, for QFTs, all possible gates would be given by bounded operators (and properly normalized).

\item A second possibility is trying to define cost functions $\tilde F$ such that they satisfy the $\mathcal{G}$-bounding property with respect to any reasonable choice of universal-gates set $\mathcal{G}$. 
The most obvious way of constructing such cost functions seems to be including an overall factor which divides a given $F$ by the maximum value taken by this cost when evaluated on the given set of universal gates, namely,
\begin{equation}\label{Fmax}
\tilde{F}[U,H]\equiv \frac{F[U,H]}{F_{\rm sup}[h]}\, , \quad \text{where} \quad F_{\rm sup}[h]\equiv \underset{e^{-ih_{(j)}}\in\mathcal{G}}{{\rm sup}}  F[U,h_{(j)}]\, .
\end{equation}
In the case of state-dependent measures, the cost function explicitly depends on the value of $U$, so $F_{\rm sup}[h]$ would need to be maximized also over the possible values of $U$.\footnote{One could also consider that the maximization occurs only with respect to the values of $U(s)$ along the minimal curve, \ie we would choose the maximum value taken by $F[U(s),h_{(j)}]$ understood as a function of $j$ and the affine parameter $s$ along the minimum trajectory. However, this prescription would produce different cost functions for different protocols, which does not seem ideal. } Another natural possibility is to normalize $F$ by the norm of the $h_{(j)}$ whose norm is maximal, namely\footnote{This would work for homogeneous costs. For inhomogeneous costs, one could in principle consider a straightforward modification of \req{Fmax2}.}
\begin{equation}\label{Fmax2}
F_{\rm sup}[h]\equiv \underset{e^{-ih_{(j)}}\in\mathcal{G}}{{\rm sup}}   || h_{(j)} ||\, .
\end{equation}
Note that with these definitions, $\tilde{F}$ actually depends on  $\mathcal{G}$, which is the price to pay for defining it in a way such that it is $\mathcal{G}$-bounding for any $\mathcal{G}$. Again, in the QFT context, the gates in $\mathcal{G}$ must always correspond to bounded operators, otherwise we would have $F_{\rm sup}=\infty$.
\end{enumerate}

An important issue which seems to have been deliberately ignored in the literature concerns the definition of reasonable universal gate sets beyond qubit systems. In that context, the set of single- and two-qubit gates is universal, providing natural notions of ``simple'' or ``small'' unitaries which can be used to build arbitrary ``big'' $n$-qubit unitaries. From our more general perspective here, $\mathcal{G}$ is a subset of $\mathcal{M}$ and the universality condition can be though of as asking $\mathcal{G}$ to be such that by multiplication of its elements we can reach a dense subset of the full unitaries manifold $\mathcal{M}$.\footnote{One could also relax this condition slightly by introducing some finite tolerance on the precision we ask from  $\mathcal{G}$.} However, the problem of finding examples of universal sets becomes quite challenging as soon as we move from the case of discrete degrees of freedom. The way this has (not) been tackled so far involves 
declaring that the geometric approach of computing geodesics on $\mathcal{M}$ provides ---somewhat by definition--- meaningful notions of ``complexity''. At that level, this just means that we define some quantity (length of geodesics on $\mathcal{M}$), we give it  the same name as the complexity defined at the beginning of Section \ref{bounds}, and then we hope that both things are somehow related. As we can see, the lower-bound relation existing between the two in the qubit-systems context can be formally extended to the general case, even though providing explicit examples of universal gate-sets even in a priori simple systems such as free quantum field seems challenging. 

Regarding this issue, we would like to make some further observations. 
As we review in Section \ref{symmm}, the typical sets of gates considered in the QFT complexity literature correspond to symmetry transformations ---\ie they are such that the generators $h_j$ form a Lie algebra, making it possible, in particular, to perform the sum of nested commutators appearing in \req{nested}. This is not extremely satisfactory. The reason is that an algebraic structure closes into itself, whereas in the complexity scenario we would like to start from a subset which precisely has the opposite property, namely, it should be such that multiplication of its elements moves us out of the original set.\footnote{The paradigmatic case is that of two-qubit gates: when we commute generalized Pauli matrices involving two-qubit gates (\ie with Pauli matrices instead of identities in two of the slots), the result generically produces a three-qubit gate, and so on.} Naturally, this is related to the universal-set issue described above, in the sense that considering certain group generators as our set of gates we are only capable of reaching unitaries belonging to a subset of $\mathcal{M}$ corresponding to the group generated by them.

 In spite of this, we would like to stress that the $\mathcal{G}$-bounding condition still makes sense when we restrict the analysis to a subset of $\mathcal{M}$ ---namely, to the one really generated by the corresponding $\mathcal{G}$. From that perspective, we can still define notions of  $\mathcal{G}$-bounding  metrics which provide lower bounds to the ``complexity'', understood now as a quantity computable only for unitaries reachable by $\mathcal{G}$ (and with respect to gates exclusively belonging to $\mathcal{G}$).

While we will not pursue the construction of explicit universal sets $\mathcal{G}$ for continuous systems  here, we think this would be a relevant and interesting problem. Quantum harmonic oscillators on a circle or free fermions on a lattice look like reasonable systems where one could try to start tackling this problem.

\section{Discriminating cost functions}\label{discri}
In the previous section we introduced a plethora of possible cost functions and we examined the $\mathcal{G}$-bounding conditions under which they provide trustable lower bounds to the quantum complexity associated to different sets of gates. In this section we establish a hierarchy among the different cost functions. The working principle consists in identifying those metrics which provide the tightest possible lower bounds. In particular, we show that inhomogeneous costs such as $F_{\kappa}$ or $F_{\sigma^2}$ must be discarded as candidate complexity measures: on the one hand, they do not provide lower bounds for complexity; on the other, they significantly violate Lloyd's bound in simple situations. In this regard, we argue that previous attempts at utilizing this kind of costs were based on confussions regarding when to expect volumetric scalings for complexity and when, on the other hand, it is natural to expect $\mathcal{O} (V^{1/2})$ scalings.  Among the possible metrics introduced in the previous section, our analysis will select $F_{\braket{H^2}}$, $F_{||H||}$ and perhaps $F_{{\rm Tr} H^2}$ as the most reasonable candidates. Finally, in section \ref{penaltyyexpulsion} we will deal with the issue of penalty factors. We will argue that our criterion of finding the tightest possible cost measures naturally leads to assigning infinite costs to all directions not belonging to the universal gate set (regardless of the choice of cost function) and how this can be implemented through the use of Lagrange multipliers.

\subsection{A no-go argument for inhomogeneous costs}
Let us analyze in more detail  the condition imposed by \req{rss}. As we saw above, fixing the integral of $r(s)$ in each interval to $1/N_{\mathcal{G}}$ was required in order for \req{proof} to go through, implying a lower bound for complexity  for the corresponding cost function.  We can understand this choice from a different perspective, as follows. In the smoothed protocol considered above ---defined by the instantaneous Hamiltonian $H_1(s)\equiv r(s) H(s)$ where $H(s)=N_{\mathcal{G}}h_j$ for $(j-1)/N_{\mathcal{G}}< s < j/N_{\mathcal{G}}$--- we can write our smooth piecewise path explicitly as
\begin{equation}
V(s)=e^{-i r(s) N_{\mathcal{G}} \left(s-\frac{j-1}{N_{\mathcal{G}}}\right)h_{(j)}}e^{-i h_{(j-1)}}\cdots e^{-ih_{(1)}}\mathds{1}\,  \quad \text{for} \quad \frac{j-1}{N_{\mathcal{G}}}< s < \frac{j}{N_{\mathcal{G}}}\, .
\end{equation}
Since in each interval the instantaneous Hamiltonian is given by $r(s) N_{\mathcal{G}} h_{(j)}$, we can alternatively use \req{usp} to write the evolution from the beginning of some interval to its final point as a path-ordered integral  of the form
\begin{align}
V(b)&=\mathcal{P} e^{-i \int_{a}^{b} r(s) N_{\mathcal{G}} h_{(j)}ds }  V(a)\\ \notag &=\mathcal{P}  e^{-i \int_{a}^{b} r(s) N_{\mathcal{G}} h_{(j)} ds}  e^{-i h_{(j-1)}}\cdots e^{-ih_{(1)}}\mathds{1}\, ,
\end{align}
where we used the notation $a\equiv (j-1)/N_{\mathcal{G}}$ and $b \equiv j/N_{\mathcal{G}}$.
But now, since $h_{(j)}$ commutes with itself, we can simply write
\begin{equation}\label{eres}
V(b)= e^{-i h_{(j)}  N_{\mathcal{G}} \int_{a}^{b} r(s) ds }    e^{-i h_{(j-1)}}\cdots e^{-ih_{(1)}}\mathds{1}\, .
\end{equation}
Naturally, when $s$ reaches the end of the interval, $s=b$, we need the gate  $e^{-i h_{(j)}}$ to have been fully implemented, in other words,
\begin{equation}
V(j/N_{\mathcal{G}})= e^{-i h_{(j)} }    e^{-i h_{(j-1)}}\cdots e^{-ih_{(1)}}\mathds{1}\, .
\end{equation}
Comparing this expression with \req{eres}, it immediately follows that condition (\ref{rss}) needs to be imposed. Otherwise, the unitary we would be considering would not correspond to the one resulting from the product of the gates.

Bearing this in mind, we can revisit the proof of the lower bound in the previous subsection in the case of a cost function $F$ which is not positive homogeneous of degree one, but rather of degree $p$, namely,
\begin{equation}
F[U(s),\alpha H(s)]=\alpha^p F[U(s), H(s)] \quad \text{for any} \quad \alpha \in  \mathbb{R}^+\, ,
\end{equation}
and any $H(s)$. This is the case of $F_{\kappa}$ and $F_{\sigma^2}$ introduced above, for which $p=\kappa$ and $p=2$ respectively.  The analogous version of \req{proof} for inhomogeneous costs of degree $p$ reads 
\begin{equation}
C_F=\int_0^1 ds F[V(s),H_1(s)]=\int_0^1  r(s)^p F[V(s),H(s)] ds \leq N_{\mathcal{G}}^p \int_0^1 r(s)^p  ds \,  .
\end{equation}
What can we say about the last integral? On the one hand, \req{rss} imposes $\int_0^1 r(s)ds=1$, which does not depend on $ N_{\mathcal{G}}$. Hence, the result for the above integral will be given by
\begin{equation}
\int_0^1 r(s)^p  ds= f(p)\, ,
\end{equation}
for some function $ f(p)$ which does not depend on $N_{\mathcal{G}}$ either. This is somewhat natural: the integral of $r(s)$ over the full path should not depend on how many gates we are using in our circuit, but only on the specific form of the chosen function. Hence, for this kind of cost functions we find
\begin{equation}\label{dps}
d_F(U_f) \leq N_{\mathcal{G}}^p\,   f(p)\, .
\end{equation}
If we wanted the geodesic distance associated to the cost $F$ to generically provide a lower bound for the minimum number of gates,  $N_{\mathcal{G}}$,  we would need  $ f(p) \sim N_{\mathcal{G}}^{1-p}$. However, as we said above, $f(p)$ has no scaling with $N_{\mathcal{G}}$, so this is not possible. The inhomogeneous costs $F_{\kappa}$ and $F_{\sigma^2}$ were attempts to increase the power of the scaling with volume of the computational costs in different contexts. By choosing, say, $p=2$, the corresponding calculations of $d_F(U_f)$ give a bound on the square of  complexity, but say nothing about the complexity itself and, in particular, they can exceed it arbitrarily. Indeed, we will see in a moment that the use of inhomogeneous costs would also give rise to crude violations of Lloyd-like bounds \cite{2000Natur.406.1047L}.   The outcome of this discussion is that inhomogeneous costs are generally unable to provide reasonable bounds for complexity, and therefore should not be considered as such.

\subsection{Metric hierarchies in the gate set $\mathcal{G}$}\label{hierar}

Previously we have analyzed which complexity metrics provide real lower bounds. Although this analysis has ruled out some possibilities, there is still a large zoo of complexity metrics at our disposal. We would like a principle which helps us to choose among them. In this regard, our proposal is the following. Since the minimal geodesics of all properly constructed local metrics are lower bounds to the circuit complexity, we need to select the one which provides the tightest lower bound, \ie the maximal one. To achieve this, we just order local metrics according to the distances they assign to equal protocols, establishing a hierarchical structure. The greater the distance, the tighter the bound.

How do we construct the tightest $\mathcal{G}$-bounding measure? The answer is divided into two parts, regarding the behavior of the local metric when evaluated on elements of the gate set $\mathcal{G}$, and on elements outside it. In this section we comment on the hierarchy with respect to the behavior on the gate set.

Among the basis-independent and homogeneous costs, it is possible to establish the following hierarchy chain:
\begin{equation}\label{hier}
F_{||H ||}\geqslant F_{\braket{H^2}}\geqslant F_{\rm FS}\;.
\end{equation}  
Notice that this is possible because the overall normalization in all three metrics can be chosen to be same, namely, the inverse of the maximum norm over the full gate set ---see \req{Fmax2} above. Apart from that observations, the first inequality is obvious, since $||H ||$ is defined as the maximum over all possible brackets $\braket{H^2}$. The second inequality is also obvious, since
\begin{equation}
F_{\rm FS}^2=F^2_{\braket{H^2}}-\braket{H}^{2}\leq F^2_{\braket{H^2}}\;,
\end{equation}
where we have omitted the overall normalization. 

Extending the arguments of \cite{Caputa:2018kdj}, we can estimate the differences between $F_{\rm FS}$ and $F_{\braket{H^2}}$  in different physical situations.
Consider for instance a system with a large mean energy $E$. In such scenario, statistical mechanics expectations are:
\begin{equation}
\braket{H^2}\simeq E^{2}\;,
\end{equation}
while 
\begin{equation}
\braket{H^2}-\braket{H}^{2}\simeq T E\;,
\end{equation}
where $T$ is the temperature of the system, which is an intensive quantity. Therefore, the cost function $F_{\braket{H^2}}$ gives a much tighter lower bound than the Fubini-Study metric, and indeed one that scales differently with the size of the system, since the energy will typically scale as the volume.\footnote{We discuss this in more detail below, and include also examples in which such scaling is different.} Indeed, the origin of these inequivalent scalings can be simply described in a situation in which the instantaneous Hamiltonian of an $n$-party system is constant and given by a sum of tensor products of the form
\begin{equation}
H=
 H_1\otimes \mathds{1}\otimes \dots \otimes\mathds{1} + \mathds{1}\otimes H_2\otimes \mathds{1}\otimes \dots \otimes\mathds{1} +\dots+\mathds{1}\otimes \dots \otimes \mathds{1}\otimes H_n  \, .
\end{equation}
 For an initially factorized state $\ket{\psi(0)}\equiv \ket{0}_1\otimes \dots \otimes \ket{0}_n$ one then finds
\begin{align} \label{h22}
\braket{\psi(s)| H(s) |\psi(s)}&=\sum_i^n \braket{\psi_{i}(s)| H_i |\psi_{i}(s)} \, , \\ \notag
\braket{\psi(s)| H(s)^2 |\psi(s)}&= \sum_i^n  \braket{\psi_{i}(s)| H^2_i |\psi_{i}(s)}+ \sum_{i\neq j}^n  \braket{\psi_{i}(s)| H_i |\psi_{i}(s)}\braket{\psi_{i}(s)| H_j |\psi_{i}(s)} \, .
\end{align}
We observe that the energy is proportional to $n$, namely, $\braket{H} \sim n$, whereas $\braket{H^2}\sim n^2$ because of the second term in \req{h22}. Both features occur as long as $\braket{\psi_{i}(s)| H_i |\psi_{i}(s)}\neq 0$. On the other hand, we have
\begin{align}
\braket{H^2}-\braket{H}^2& =\sum_i \left [ \braket{\psi_{i}(s)| H^2_i |\psi_{i}(s)}- \braket{\psi_{i}(s)| H_i |\psi_{i}(s)}^2\right] \sim n\, .
\end{align}
Therefore, we observe that while $F_{\braket{H^2}}=\sqrt{\braket{H^2}}\sim n \sim E$, the Fubini-Study one yields   $F_{\rm FS}=\sqrt{\braket{H^2}-\braket{H}^2}\sim \sqrt{n} \sim \sqrt{E}$. An explicit realization of this phenomenon, in a context of direct physical importance, was found in \cite{Caputa:2018kdj} for Virasoro protocols in two-dimensional CFTs. In that case, $F_{\braket{H^2}}$ scales with the central charge $c$ whereas $F_{\rm FS}\sim \sqrt{c}$. We review this below in Section \ref{vira}.

A similar phenomenon to the one just explained occurs when one considers the basis-dependent cost $F_2$ and the trace cost $F_{{\rm Tr}H^2}$. The reason is as follows. We showed in \req{trH2} that $F_{{\rm Tr}H^2}$ actually agrees with $F_2$ whenever the basis of Hermitian operators in which we expand the instantaneous Hamiltonians satisfies ${\rm Tr} [K_I K_J]=\delta_{IJ}$. As we have said, $F_{{\rm Tr}H^2}$ is nothing but $F_{\braket{H^2}}$ in a particular state (the maximally mixed one), so one could naively think that using $F_2$ (or $F_1$) one should get the same scalings as those expected for $F_{\braket{H^2}}$ rather than the ones expected for $F_{\rm FS}$. This is not the case, however. The reason is that if the basis of Hermitian operators contains $I=1,\dots,M$ generators, ${\rm Tr} [K_I K_J]$ will be a $M\times M$ matrix, and generically
\begin{equation}
F_{{\rm Tr}H^2}=\sqrt{\sum_{I,J} Y^I(s)Y^J(s) {\rm Tr}[K_I K_J]}\sim \sqrt{M^2}\sim M\, ,
\end{equation}
whereas 
\begin{equation}
F_2 =\sqrt{\sum_I \left( Y^I(s)\right)^2} \sim \sqrt{M}\, ,
\end{equation}
namely, the scaling will be different in general. This will change in the case that  ${\rm Tr} [K_I K_J]=\delta_{IJ}$, which effectively reduces the $M\times M$ matrix to the $M$ elements of its diagonal. Saying it differently, whenever we have an orthonormal basis we have ${\rm Tr} [K_I ]=0$, since one of the generators is the identity. Only in these cases, $F_{{\rm Tr}H^2}=F_2 \sim \sqrt{M}$. This illustrates the fact that using the basis-dependent cost functions $F_2$ (or equivalently to $F_1$) is in general inequivalent to using $F_{{\rm Tr}H^2}$. In a certain sense,  $F_2$ is a ``Fubini-Study-like'' version of $F_{{\rm Tr}H^2}$, and therefore provides worse lower bounds to the circuit complexity in general. The fact that using the homogeneous costs $F_2$ and $F_{\rm FS}$ one is led to lower bounds on complexity which scale with the square root of the volume instead of the volume itself ---as observed in different contexts \eg \cite{Jefferson:2017sdb,Chapman:2017rqy,Hackl:2018ptj,Guo:2018kzl,Belin:2018bpg}--- has the simple origin just described.
 

\subsubsection{Inhomogeneous metrics violate Lloyd's bound} \label{vira}
In many contexts in which an apparently incorrect scaling was found, inhomogeneous definitions of complexity were proposed to deal with this issue. From our perspective, such notions are not only \emph{ad hoc}, but actually fail to provide lower bounds to the quantum complexity. 
In addition to this, inhomogeneous metrics can be seen to violate Lloyd's bound in several situations, as we show here.

The reason of the violation is simple. As it has been argued above, generically, Fubini-Study and related choices scale as $F_{\rm FS}\propto \sqrt{E}$. Squaring and integrating over a trajectory provides a cost that ``saturates'' Lloyd's bound $C_{\sigma^{2}}\sim \int F_{\rm FS}^{2}dt=\int F_{\sigma^2}dt \propto Et$. Parametrically, the regime when this occurs is characterized by $\langle E^{2}\rangle\simeq \langle E\rangle^{2}$. The problem happens then in the opposite regime, whenever we have a situation in which $\langle E^{2}\rangle\gg \langle E\rangle^{2}$. In these cases we have:
\begin{equation}
F_{\langle H^{2}\rangle}\simeq F_{\rm FS}\, , \quad \text{and thus}\quad F_{\sigma^{2}}=F_{\rm FS}^{2}\simeq F_{\langle H^{2}\rangle}^{2}\, .
\end{equation}
But in these cases, the typical energies involved in the process are of order $E\sim \sqrt{\sigma_{H}^{2}}=F_{\rm FS}\simeq F_{\langle H^{2}\rangle}$. Therefore, squaring Fubini-Study to get nice results in the typical high energy regimes will destroy the consistency of the approach in other regimes by violating Lloyd's bound.

But if one cannot square the cost, how are bulk volumes obtained in the holographic context? This was explained within the Path-integral complexity proposal \cite{Caputa2017}, and goes as follows. In Poincar\'e coordinates, a constant time slice of the bulk metric is given by 
\begin{equation}\label{timeslice}
ds^{2}=\frac{dz^{2}+dx_{\perp}^{2}}{z^{2}}\, .
\end{equation}
Following the usual ideology of the AdS/MERA correspondence \cite{PhysRevD.86.065007}, we should see this metric as a circuit in the $z$-coordinate that generates the vacuum at the boundary. This circuit is necessarily time-dependent, where ``time'' is now the radial $z$-coordinate itself. Within the Path-Integral proposal, this ``time'' is basically the Euclidean time in the Liouville action, and the solutions corresponding to AdS are obviously $z$-dependent. In turn, this non-trivial $z$-dependence is required to obtain the right scaling with the volume.

So if we are to obtain bulk volumes (and geometry) from conformal transformations, we should find the (Euclidean) time-dependent transformations, such that the geometry at each $z$ is~(\ref{timeslice}). Falling beyond the scope of the present paper, we leave specific calculations for future work, but it is still illustrative to see how the different 2d-CFT complexity metrics behave in ``time-dependent'' scenarios. Let us then regard the radial coordinate $z$ as a protocol time $s$. We want a circuit that generates at time $s$ a conformal transformation (a diffeomorphism) $[f(s,x_{+}),f(s,x_{-})]$ such that the metric at time $s$ is given by~(\ref{timeslice}). This circuit is built up from infinitesimal diffeomorphisms $\epsilon (s,x_{+})$ and $\epsilon (s,x_{-})$ which, when composed all together, generate $[f(s,x_{+}),f(s,x_{-})]$.\footnote{See Appendix~\ref{ap1} and \cite{Caputa2017} for a more detailed discussion of conformal circuits.} Since conformal transformations are represented in the Hilbert space by unitaries and infinitesimal conformal transformations are generated by the stress tensor, the circuit would read\footnote{We only consider left movers. The discussion extends trivially to $x_{-}$.}
\begin{equation}\label{usp2}
U_{f(s,x_{+})}=\mathcal{P} e^{-i \int_{0}^s (\int\epsilon (s',x_{+})T_{++}(x^{+})dx^{+})ds'}\, .
\end{equation}
As usual, we first need to relate the instantaneous Hamiltonian ---in this case parametrized by $\epsilon (s,x_{+})$--- to the finite conformal transformation $f(s,x_{+})$ ---see Appendix~\ref{ap1}. For the present purposes, we only need to notice that such instantaneous Hamiltonian $H(s)=\int \epsilon (s,x_{+})T_{++}(x^{+})dx^{+}$ will be time dependent, in order to generate the $z$ dependent metric~(\ref{timeslice}). The initial state $\vert \psi (0)\rangle$ can be the vacuum or any other state, depending on the situation. In any case, the state at time $s$ is a conformal transformation of the initial one $\vert\psi (s)\rangle = U_{f(s,x_{+})}\vert \psi (0)\rangle$. Before computing $F_{\langle H^{2}\rangle}$ or $F_{\rm FS}$, let us start with
\begin{equation}
\braket{\psi(s)| H(s) |\psi(s)} =\langle\psi (0)\vert U_{f(s,x_{+})}^{\dagger}\left(  \int \epsilon (s,x_{+})T_{++}(s,x^{+})dx^{+}\right)  U_{f(s,x_{+})}\vert \psi (0)\rangle\;.
\end{equation}
Crucially, this can be obtained using the known behavior of the stress tensor under conformal transformations,
\begin{equation}
 U_{f(s,x_{+})}^{\dagger}T_{++}(s,x^{+}) U_{f(s,x_{+})}=f'(s,x^{+})^{-2}\left(T_{++}(s,x^{+})-\frac{c}{12}\lbrace f(s,x_{+}),x_{+}\rbrace \right)\;,
\end{equation}
where $\lbrace f(s,x_{+}),x_{+}\rbrace$ is the Schwarzian derivative. One simply obtains
\begin{equation}
\braket{\psi(s)| H(s) |\psi(s)}=\int  \frac{ \epsilon (s,x_{+})}{f'(s,x^{+})^{2}}  \left(\langle T_{++}(s,x^{+})\rangle-\frac{c}{12}\lbrace f(s,x_{+}),x_{+}\rbrace \right) dx^+\;.
\end{equation}
Hence, if $\braket{\psi(s)| H(s) |\psi(s)}$ is to be $\mathcal{O}(c)$ ---and therefore $\mathcal{O}(V)$ in applications to holography--- we have two possibilities. Either we start with a state with scaling dimension of order  $\mathcal{O}(c)$, so that the first term in the parenthesis is of that order, or we consider a conformal transformation so that the Schwarzian term becomes effective. In the second case, in the limit $c\rightarrow \infty$ ---which is the one interesting for holographic applications--- we have 
\begin{equation}
\braket{\psi(s)| H(s) |\psi(s)}\xrightarrow [c\rightarrow\infty]{}-\int  \frac{ \epsilon (s,x_{+})}{f'(s,x^{+})^{2}}\frac{c}{12}\lbrace f(s,x_{+}),x_{+}\rbrace dx^+ \;.
\end{equation}
Moreover, in the same limit it is simple to observe that
\begin{equation}
F_{\langle H^{2}\rangle}\xrightarrow [c\rightarrow\infty]{}\vert\int  \frac{ \epsilon (s,x_{+})}{f'(s,x^{+})^{2}}\frac{c}{12}\lbrace f(s,x_{+}),x_{+}\rbrace dx^+ \vert\;.
\end{equation}
The conformal transformation $U_{f(s,x_{+})}$ required to have the ``time-dependent'' metric~(\ref{timeslice}) is such that the Schwarzian term contributes\footnote{This is because the conformal transformation cannot be part of the global subgroup of the conformal group, since such subgroup is precisely the one that leaves invariant the vacuum.} (it is even dominant near the vacuum), and $F_{\langle H^{2}\rangle}$ is forced to be $\mathcal{O}(c)$, and will be sensitive to the bulk volume. This is exactly what happens within the Path-integral complexity proposal \cite{Caputa2017}, where we just insert the solution~(\ref{timeslice}) into the Liouville action, which in turn can be written as the Schwarzian action \cite{Alekseev:1988ce,Caputa:2018kdj}.  On the other hand, the Fubini-Study choice is not sensitive to the Schwarzian, since this is a constant which disappears when subtracting the one-point function in the definition of the Fubini-Study metric. Fubini-Study is then $\mathcal{O}(\sqrt{c})$. For a precise computation of the Fubini-Study metric for Virasoro circuits see \cite{Caputa2017}.

Summarizing, $F_{\langle H^{2}\rangle }$ is a true lower bound to complexity. It also shows the right scaling in the appropriate cases. On the other hand, Fubini-Study provides suboptimal scalings in simple scenarios. Finally, the ``squaring'' solution is not a lower bound, and although it might get the right scaling at high energies, it typically does not  do so at low energies, where it violates Lloyd's bound \cite{2000Natur.406.1047L}. 

Finally, let us also comment on $F_{{\rm Tr}H^2}$. As we have mentioned, this can be thought of as $F_{\braket{H^2}}$ where the instantaneous state $\ket{\psi(s)}\bra{\psi(s)}$ has been replaced by the maximally mixed one, $\rho_{\rm mixed}=\mathds{1}/{\rm dim} \mathcal{H}$. This is nothing but a thermal state at infinite temperature, $\rho_{\rm mixed}= e^{-\beta H}/{\rm dim} \mathcal{H}$ with $\beta=0$, so  $F_{{\rm Tr}H^2}\sim E$ at high (order of the cutoff) energies. Since by definition $F_{{\rm Tr}H^2}$ does not depend on the actual state of the system, this suggests that this measure could potentially give rise to violations of the Lloyd bound at low energies. To fix this issue, one can consider such metric in appropriate microcanonical scenarios. This can be implemented through the use of Lagrange multipliers, constraining the energy to be fixed to some specific value ---see \eg \cite{2015PhRvL.114q0501W}. We discuss these constraints in greater generality in the next section. At any rate, in such situation, $F_{{\rm Tr}H^2}$ would become a state dependent metric, very much alike $F_{\braket{H^2}}$. Similar comments apply to the state-independent measure $F_{||H||}$.

\subsection{Metric hierarchies outside $\mathcal{G}$: Penalties as constraints}\label{penaltyyexpulsion}
So far we have said nothing about the possibility of penalizing gates (or directions in $\mathcal{M}$) which may be ``more expensive'' than others. 
A priori, a general discussion about penalties might seem slightly arbitrary, and this has indeed been the case in recent literature ---see \eg \cite{2005quant.ph..2070N,Brown:2017jil,Magan:2018nmu} for several approaches. The goal of this subsection is to argue that the previous discussion about the tightest bound suggests a somewhat canonical way of approaching this problem. Just like when discussing cost-function hierarchies within the gate set $\mathcal{G}$, the guiding principle outside the gate set $\mathcal{G}$ is to ensure that the resulting metrics provide the tightest possible lower bounds. Let us describe how to achieve this feature.

Consider a gate-set $\mathcal{G}$ composed of  unitaries $U_{I}=e^{-i K_{I}}$. We can expand a generic instantaneous Hamiltonian as
\begin{equation}
H(s)=\sum\limits_{e^{-i K_I} \in \mathcal{G}}Y_I(s) K_I + \sum\limits_{e^{-i K_I'}\not\in \mathcal{G}}Y'_I(s) K_I'\, .
\end{equation} 
The conditions for a metric $F$ to be a lower bound are the $\mathcal{G}$-bounding ones appearing in \req{gbounding}, namely,
\begin{equation}
F(e^{-i K_I}\in\mathcal{G})\leq 1\;.
\end{equation}
To find the metrics providing the tightest lower bounds we should maximize over the space of metrics satisfying the $\mathcal{G}$-bounding constraint. Then, if possible, we should choose costs saturating the $\mathcal{G}$-bounding constraint for $e^{-i K_I}\in\mathcal{G}$, namely, metrics such that
\begin{equation}\label{consr}
F(e^{-i K_I}\in\mathcal{G})=1\;.
\end{equation}
Examples of cost functions satisfying this property in the context of $n$-qubit systems are $F_{{\rm Tr}H^2}$, $F_{\braket{H^2}}$ and $F_{||H ||}$. For those, one has
\begin{equation}\label{metrics}
F_{{\rm Tr}H^2}(\sigma\in\mathcal{G})=1\, , \quad F_{\braket{H^2}}(\sigma\in\mathcal{G})=\sqrt{\langle \psi (\tau)\vert \sigma^{2}\vert\psi (\tau)\rangle }=1\, , \quad F_{||H ||}(\sigma\in\mathcal{G}) =1\, ,
\end{equation}
which follows from the fact that the  generalized spin operators built from tensor products of Pauli matrices and the identity ---an example being $
\sigma=\sigma_{x}\otimes\sigma_{y}\otimes\mathds{1}\otimes\cdots\otimes\sigma_{x}
$--- can be normalized so that they satisfy ${\rm Tr} (\sigma_I \sigma_J)=\delta_{IJ}$ or $\sigma_I^2=\mathds{1}$.
Note that in that context, \req{consr} would not be compatible \eg with the Fubini-Study cost, since such metric typically does not give cost one to the $\sigma_I$.

Now, since the $\mathcal{G}$-bounding conditions do not constrain gates $e^{-i K_I'} \not\in \mathcal{G}$, we should assign as high as possible  costs to those. In fact, we should actually assign infinite costs to them in order to maximize the distance of any possible circuit in the complexity geometry. We emphasize that this does not imply that distances between points go to infinity, since if $\mathcal{G}$ is a universal gate set, we can always find trajectories from any point to any other with finite distance. Trajectories would just need to be constructed from gates involving arbitrary superpositions of $e^{-i K_I}\in\mathcal{G}$. In fact, as shown in \cite{2006Sci...311.1133N} in the case of $n$-qubit systems ---where $\mathcal{G}$ contains 1- and 2-qubit operators--- including sufficiently high penalties associated to the directions not belonging to $\mathcal{G}$ is basically the same as including infinite penalties, the error scaling with the inverse of the penalty size. It is natural to expect this equivalence to hold in general. Giving infinite cost to non-$\mathcal{G}$ directions is a natural consequence of looking for the tightest lower bound.

The question now is how to implement this condition. We have two possibilities. The first is to work with metrics including penalties. For example, Nielsen proposed \cite{2005quant.ph..2070N}
\begin{equation}\label{Npen}
F_{1p}\equiv \sum\limits_{\sigma}p(\textrm{wt}(\sigma))\,\vert c_{\sigma}\vert\, , \quad
F_{2q}\equiv \sqrt{\sum\limits_{\sigma} q(\textrm{wt}(\sigma))\,(c_{\sigma})^{2}} \;,
\end{equation}
as generalizations of $F_1$ and $F_2$, where $p(\textrm{wt}(\sigma))$ and $q(\textrm{wt}(\sigma))$ are the penalty functions associated to the gate $\sigma$ and $\textrm{wt}(\sigma)$ is the number of single-site Pauli matrices (not identities) appearing in the tensor product of $\sigma$. Then, one would solve the geodesic equation in the limit in which penalties go to infinity. 
This seems a bit cumbersome in general. 
Notice that in such a limit, the solutions to the geodesic equation $U(s)$ are such that
\begin{equation}\label{constraint}
iU^{-1}(s)\dot U(s)=\sum\limits_{e^{-i K_I} \in \mathcal{G}}Y_I(s) K_I \;,
\end{equation}
with no contributions from gates outside the universal set. So another way to implement the condition is to choose a metric satisfying \req{consr} for the whole complexity geometry, and solve the geodesic equation with \req{constraint} imposed as a constraint. This can be  done using Lagrange multipliers $\lambda_I'$. The final geometric-complexity functional would be given by
\begin{equation}
C_F=\int_0^1 ds \left[F[U(s),H(s)]+\sum\limits_{e^{-i K_I'}  \not\in \mathcal{G}}\lambda_{I}' {\rm Tr} \left[K_I' H(s) \right] \right]\, .
\end{equation}
This seems to be, in particular, the most efficient way of imposing $k$-locality, which is the statement that we can only use generalized Pauli matrices with weight $w(\sigma)\leq k$. In such scenario, our gate set $\mathcal{G}$ is chosen to contain all generalized Pauli matrices with weight $w(\sigma)\leq k$ and taking the limit of infinite penalties amounts to constrain the space of trajectories to those satisfying $k$-locality. 
 Observe that in this derivation we have used the possibility of imposing \req{consr}, which relied on the orthogonality of the basis operators. More generally, constraints will be specific functionals of the instantaneous gate, so that
\begin{equation}
C_F=\int_0^1 ds \left[F[U(s),H(s)]+\sum\limits_{i  }\lambda_{I} f_{i}(H(s)) \right]\, . 
\end{equation}

In the QFT context it is generally less clear how to choose $\mathcal{G}$ and therefore which directions should be assigned infinite costs. For CFTs, it was argued in \cite{Magan:2018nmu} that the quantity playing the role of $\textrm{wt}(\sigma)$ should be the scaling dimension $\Delta$ of the associated field. The reason is transparent for large-$N$ theories, where such scaling dimension is additive with respect to multiplication of field operators, as happens with $\textrm{wt}(\sigma)$ for spin systems. Another reason appears when thinking about the operator product expansion of fields, which raises two observations. First, it is clear that the ``weight'' is not an intrinsic property of the operator, since the OPE ensures that it is enough to consider local operators as quantum gates, as long as we consider all of them. Second, it is clear that multiplying many low-dimension operators increases the number of high-dimension operators in the OPE. It is thus natural to penalize operators with high scaling dimension. This is exactly what we expect in the black hole context, where black-hole-creating operators should be strongly penalized. Then, in the CFT context, it is natural to impose the condition that we can only use fields with scaling dimension $\Delta \leq \Delta_{c}$.\footnote{Notice that, strictly speaking, this condition implies strict local gates in the CFT, since any non-local interaction would generically contain in its OPE operators of any desired dimension. But it is also the condition that ensures $k$-locality in the internal color space when the CFT is a large-$N$ gauge theory.} In this context, the string of Lagrange multipliers would be associated to fields with dimensions $\Delta>\Delta_{c}$.

On the other hand, as we mentioned above, the gate sets $\mathcal{G}$ usually considered in the QFT context are not really universal. Rather, they generate only a subgroup of the unitary group $\mathcal{M}$.
In this context, looking for the tightest bound, one is again led to set infinite costs to any instantaneous Hamiltonian with support outside $\mathcal{G}$. This is then directly achieved by using the same functional associated to the smaller group $\mathcal{G}$, \emph{without constraints},
\begin{equation}\label{consi}
C_{F}=\int_{\mathcal{M}} ds \left[F[U(s),H(s)]+\textrm{Constraints} \right] =\int_{\mathcal{G}} ds F[U(s),H(s)]\;.
\end{equation}
We already mentioned an important example of this situation above, which is the case of gate-sets corresponding to generators of a certain Lie algebra ---see appendix ~\ref{symmm}. 
Another example occurs when one considers the so-called generalized coherent groups \cite{perelomov1972}. These generate the set of generalized coherent states of the theory, and it can be proven that generic quantum dynamics localizes in such subspace in the semiclassical limit \cite{RevModPhys.54.407}.\footnote{Such structure has been used also in the context of large-$N$ theories to prove several aspects about entanglement entropy \cite{Magan:2017udh}.}

Yet another interesting example, of great importance in the holographic context, is that of generalized free fields. By definition, generalized free fields are those satisfying Wick's factorization, which in turn can be shown to correspond to low dimension operators in the CFT.\footnote{Operators with scaling dimensions of $\mathcal{O}(c)$, with $c$ the central charge, are expected to interact strongly with the family of the energy momentum tensor, and are not expected to satisfy large-$N$ factorization.} Generalized free fields generate an approximate algebra, only truly valid in the large-$N$ limit. In this limit, if we assign infinite costs to operators of high scaling dimension, the complexity functionals would be given by \req{consi} with 
$\mathcal{G}=G_{\textrm{GFF}}$ generated by the generalized free fields.\footnote{We compute the instantaneous Hamiltonian associated to generalized free fields in appendix~\ref{ap1}.} This then shows that the quantum complexity of the dual CFT can be computed using the free bulk dynamics,  using the known one-to-one matching between low dimension operators at the boundary and fields in the bulk.

Before closing, let us mention that the idea of imposing constraints in the form of Lagrange multipliers for $3$- or higher-site qubit operators was previously entertained in various papers ---see \eg \cite{2006PhRvL..96f0503C,2007PhRvA..75d2308C,2015PhRvL.114q0501W}. The perspective in those works is very similar in technical terms to Nielsen's approach, in the sense of considering continuous circuits and extremizing over several proposed classical actions in such space. From our perspective, Nielsen's approach naturally leads to such time-optimal control proposals, when we realize we should maximize over the zoo of $\mathcal{G}$-bounding complexity metrics in order to obtain the best lower bound.

\section{Complexity equals (geometric) action}\label{geomA}
Our search for cost functions providing the tightest possible lower bounds for complexity has led us to a few candidates, namely, $F_{\braket{H^2}}$, $F_{|\braket{H}|}$, $F_{||H||}$ and perhaps $F_{{\rm Tr}H^2}$, suitably defined in QFT in microcanonical sectors. 
$F_{\braket{H^2}}$ and $F_{|\braket{H}|}$ were previously proposed in \cite{Magan:2018nmu,Caputa:2018kdj} in the context of symmetry groups ---and more generically for coherent states--- but they were not properly justified from a quantum computational of view. The outcome of our discussion  here is that, once Lagrange multipliers are included, these cost functions provide a good and tight lower bound for complexity.

In this section we want to expand on the properties of $F_{\braket{H^2}}$ and $F_{|\braket{H}|}$. 
The reason is that these cost measures turn out to satisfy a series of additional properties which make them, to some extent, canonical choices. On the one hand, $F_{\braket{H^2}}$ turns out to correspond to the canonical Hermitian metric defined on any Hilbert space. This metric is compatible with a complex structure and a symplectic form, which defines a natural K\"ahler structure in any Hilbert space $\mathcal{H}$. The existence of this K\"ahler structure, along with the fact that the unitaries manifold of $\mathcal{H}$ posseses a Lie group structure, allows one to define a somewhat canonical distance functional in $\mathcal{M}$ known as ``coadjoint orbit action'' or ``geometric action'', which is basically given by $F_{|\braket{H}|}$. $F_{\braket{H^2}}$ also reduces to the geometric action for systems of small quantum variance ---in particular, in semiclassical limits.

Finally, the interpretation of $\mathcal{H}$ as a phase space, along with the fact that quantum evolution can be understood as Hamiltonian evolution for expectation values, suggests a canonical notion of quantum action associated to any quantum Hamiltonian, simply given by the integral of the Lagrangian associated to $\braket{H}$. The equations of motions associated to this quantum action are by construction equivalent to the Schr\"odinger equation. This quantum action may be then used as a ``gauge'' of complexity measures. In particular, we discuss its intimate relation with the geometric action.\footnote{The geometric actions construction is reviewed in Appendix \ref{ap2}. The rest of the section is more or less self contained.}

\subsection{$F_{\braket{H^2}}$ from the canonical metric in Hilbert space}

Any complex Hilbert space $\mathcal{H}$ can be considered as a real vector space $\mathcal{H}_{\mathbb{R}}$ equipped with an almost complex structure $J$, which consists on a linear isomorphism of $\mathcal{H}_{\mathbb{R}}$ satisfying $J^2=-\mathds{1}$. From this perspective, it is natural to split the Hilbert metric on $\cH$ into its real and imaginary parts:
\begin{equation}
\braket{\phi| \psi}=   g(\phi,\psi)+ i\Omega(\phi,\psi)\, ,
\end{equation}
where:
\begin{equation}\label{GOm}
g(\phi,\psi)\equiv \frac{1}{2}(\braket{\phi| \psi}+\braket{\psi| \phi}) \, ,\quad \Omega(\phi,\psi)\equiv -\frac{i}{2} \left[\braket{\phi| \psi}-\braket{\psi| \phi} \right]\, .
\end{equation}
It can be shown that $g$ and $\Omega$ define, respectively, a real and positive-definite inner product and a symplectic form on $\mathcal{H}_{\mathbb{R}}$ compatible with $J$, in the sense that:
\begin{equation}
g(\phi,\psi)=\Omega(\phi,J\psi)\, , \qquad \forall\,\, \phi , \psi \in \cH_{\mathbb{R}}\, .
\end{equation}
The triple $(\mathcal{H}_{\mathbb{R}},J,\Omega)$ (note that $(J,\Omega)$ uniquely determine $g$) canonically becomes a complex K\"ahler manifold, whose K\"ahler structure we denote again by $(J,\Omega)$ for ease of notation. Therefore, every Hilbert space comes equipped with a canonical K\"ahler structure associated to it.\footnote{This is a  standard discussion in the geometric quantum mechanics literature, see \eg \cite{Kibble:1978tm,Ashtekar:1997ud}.} In particular, in this formulation the Schr\"odinger equation for a ket $\ket{\psi}$ reads:
\begin{equation}
\partial_s \ket{\psi} =- J H \ket{ \psi}\, .
\end{equation}
Given some reference state $\ket{\psi}$ the inner product corresponding to two infinitesimally evolved states $-J H_1 \psi$ and $-J H_2\psi$ is given by
\begin{equation}\label{ggg}
g(-J H_1 \psi,-J H_2 \psi)= \braket{\psi|\{ H_1,H_2\} |\psi}\, ,
\end{equation}
where $\{,\}$ is the usual anticommutator and we used the self-adjointness of $H_1$ and $H_2$. Similarly, one finds
\begin{equation}\label{Ommm}
\Omega(-J H_1 \psi,-J H_2 \psi)=- i \braket{\psi|[ H_1,H_2] |\psi}\, .
\end{equation}
Therefore, the symplectic form and the metric respectively yield the expectation values of the commutator and anticommutator of the evolution generators. In our context, ``evolution'' along different paths in the Hilbert space is generated by the corresponding instantaneous Hamiltonian $H(s)$, and it follows from \req{ggg} that
\begin{equation}
 \frac{1}{2}g(\dot  \psi(s),\dot  \psi(s))=\braket{\psi(s)| H(s)^2 |\psi(s)} \, ,
\end{equation}
which is nothing but the square of the cost function $F_{\braket{H^2}}$ defined in \req{H2Ja}. Hence, in addition to the previously discussed features, $F_{\braket{H^2}}$ is also associated to the canonical Hermitian metric defined in every Hilbert space $\mathcal{H}$. 

Both $g$ and $\Omega$ can be related to analogous canonical structures in the projective Hilbert space $\mathcal{P}$ associated to $\mathcal{H}$. Related to $\cH$ we consider two infinite-dimensional Hilbert manifolds: i) The infinite-dimensional Hilbert sphere $\mathbb{S}^{\infty} \equiv\left\{ \ket{\psi}\in \cH\backslash\left\{ 0 \right\} \,\, \vert\,\, \braket{\psi|\psi} = 1\right\}$.
ii) The infinite-dimensional complex projective space $\mathbb{C}\mathbb{P}^{\infty} = \cH\backslash\left\{ 0 \right\}/\mathbb{C}^{\ast}$, where $\mathbb{C}^{\ast}$ acts on $\cH$ through its standard diagonal action.
Associated to these manifolds we have canonical maps
\begin{equation}
\iota\colon \mathbb{S}^{\infty} \hookrightarrow   \cH\backslash\left\{ 0 \right\} \, , \qquad \pi\colon  \cH\backslash\left\{ 0 \right\} \to \mathbb{CP}^{\infty}\, ,
\end{equation}
corresponding to the canonical embedding of $\mathbb{S}^{\infty}$ and the canonical projection onto $\mathbb{C}\mathbb{P}^{\infty}$, respectively. Furthermore, composing $\pi$ and $\iota$ we obtain a smooth submersion
\begin{equation}
\pi\circ\iota\colon \mathbb{S}^{\infty} \to \mathbb{C}\mathbb{P}^{\infty}\, ,
\end{equation}
which defines a $U(1)$ bundle over $\mathbb{C}\mathbb{P}^{\infty}$, the infinite-dimensional version of the Hopf fibration. The metric $g$ on $\cH_{\mathbb{R}}$ defines by pull-back a non-degenerate metric on $\mathbb{S}^{\infty}$ which we denote by $\iota^{\ast}g$ and which, proceeding by analogy with the finite-dimensional case, corresponds with the round metric on $\mathbb{S}^{\infty}$. Given this metric on $\mathbb{S}^{\infty}$, there exists a unique metric on $\mathbb{C}\mathbb{P}^{\infty}$ making $\pi\circ\iota\colon \mathbb{S}^{\infty} \to \mathbb{C}\mathbb{P}^{\infty}$ into a Riemannian submersion. This metric can be shown to be the Fubini-Study metric $g_{\rm FS}$. Alternatively, the Fubini-Study metric on $\mathbb{C}\mathbb{P}^{\infty}$ can be defined through the $\mathbb{C}^{\ast}$-invariant metric on $ \cH\backslash\left\{ 0 \right\}$ determined by the following global K\"ahler potential:
\begin{equation}
\mathcal{K}_{{\rm FS}}(\ket{\psi}) \equiv \log(\braket{\psi|\psi})\, , \quad \ket{\psi}\in \cH\, .
\end{equation}
For comparison, note that the K\"ahler potential of the standard ``flat'' metric on $\cH$ is given by $\braket{\psi|\psi}$. A quick computation shows that the associated metric on  $ \cH\backslash\left\{ 0 \right\}$ is given by:
\begin{equation}
g^0_{{\rm FS}}\vert_{\ket{\psi}}(\ket{\delta\psi},\ket{\delta\psi})) = \frac{\braket{\delta\psi |\delta\psi}}{\braket{\psi |\psi}}-\frac{ \braket{\psi| \delta\psi} \braket{\delta\psi | \psi}}{\braket{\psi |\psi}^2}\, ,
\end{equation}
where $\ket{\delta\psi}\in T_{\ket{\psi}}(\cH\backslash\left\{ 0 \right\})$ is an element of the tangent space of $\cH\backslash\left\{ 0 \right\}$ at $\ket{\psi}$. The metric $g^0_{{\rm FS}}$ is clearly $\mathbb{C}^{\ast}$ invariant, whence it descends to the quotient, yielding the standard Fubini-Study metric on $\mathbb{C}\mathbb{P}^{\infty}$. It is easy to verify that choosing local holomorphic coordinates $g^0_{{\rm FS}}$ and $g_{{\rm FS}}$ both reduce to the familiar expression for the Fubini-Study metric.\footnote{For this, choose first holomorphic homogeneous coordinates on $ \cH\backslash\left\{ 0 \right\}$ such that
\begin{equation}
\ket{\psi} = (\hat{z}^1,\hdots, \hat{z}^i,\hdots,\hat{z}^{\infty})\, , \quad \braket{\psi |\psi} = \sum_{1}^{\infty} \vert \hat{z}^i\vert^2\, .
\end{equation}
Choosing now the open set defined by $\hat{z}^1\neq 0$ we obtain local holomorphic coordinates on $\mathbb{CP}^{\infty}$ as follows,
\begin{equation}
(z^1,\hdots, z^i,\hdots,z^{\infty}) = \left(\frac{\hat{z}^2}{\hat{z}^1},\hdots, \frac{\hat{z}^i}{\hat{z}^1},\hdots,\frac{\hat{z}^{\infty}}{\hat{z}^1}\right)\, .
\end{equation}
Implementing this coordinates in $g^0_{{\rm FS}}$ we obtain the familiar coordinate-expression for the Fubini-Study metric.} This metric, together with the canonical complex structure on $\mathbb{C}\mathbb{P}^{\infty}$ ---making $\pi$ into a holomorphic submersion--- give rise to a symplectic form $\omega$ corresponding to the standar K\"ahler structure on $\mathbb{C}\mathbb{P}^{\infty}$. Restricted to a smooth curve $\ket{\psi(s)}$ of normalized states $g^0_{{\rm FS}}$ becomes 
\begin{equation}
g^0_{{\rm FS}}\vert_{\ket{\psi(s)}}(\ket{\delta\psi},\ket{\delta\psi})) = F_{\rm FS}^2   =\left[ \braket{\psi(s) | H(s)^2 |\psi(s)}-\braket{\psi(s) | H(s) |\psi(s)}^2 \right] \, .
\end{equation}
where have used Equation \req{sch} to characterize the tangent vectors $\ket{\delta\psi}$ to curves $\ket{\psi(s)}$ satisfying the Schr\"odinger equation.

To summarize, we see that both $F_{\braket{H^2}}$ and $F_{\rm FS}$ are related to canonical metrics in $\mathcal{H}$ and $\mathcal{P}$ respectively, which are in turn related to each other as we have just explained. While the physical space of states is $\mathcal{P}$ rather than $\mathcal{H}$, which would suggest the use of $F_{\rm FS}$ above $F_{\braket{H^2}}$, we have argued above that $F_{\braket{H^2}}$ always provides better bounds for complexity. This is precisely related to the fact that $F_{\rm FS}$, being the canonical metric in $\mathcal{P}$, assigns vanishing costs to many operations which are different from the identity (all those which move the state within its ray).






\subsection{$F_{\braket{H^2}}$ and geometric actions}\label{geomacQM}
 

Coadjoint orbit actions are reviewed in fair generality in Appendix~\ref{ap2}, which should be consulted for further references and definitions. Very briefly, the idea is the following. Any Lie group $G$ acts naturally on the dual of its Lie algebra, $\mathfrak{g}^*$, through the coadjoint representation.  As explained in more detail in Appendix~\ref{ap2}, Kirillov proved that the orbits of the coadjoint action of $G$ on $\mathfrak{g}^*$ are symplectic submanifolds with respect to a canonically defined and invariant symplectic structure. Hence, each such orbit can in principle be considered as a ``classical phase space'' and, from this point of view, it is natural to consider trajectories $\psi(s) \in \mathfrak{g}^*$ satisfying physical equations. The integral along the affine parameter of the pairing of $\psi(s)$ with the Lie algebra element which produces infinitesimal motion along the trajectory ---which plays the role of instantaneous Hamiltonian in the language utilized throughout the paper--- defines the so-called ``geometric action'' or ``coadjoint orbit action'' \cite{Kiri}.

Now, every quantum-mechanical system has an associated  Hilbert space $\mathcal{H}$ and a continuous group  of unitaries  $\mathcal{M}$  which acts on it. As explained in Section \ref{sec:BLgroups}, using the exponential in the norm topology we can identify the Lie algebra of $\cM$ as $\mathfrak{u}(\cH)$, that is, the Lie algebra of all skew-Hermitian operators on $\mathcal{H}$. Hence, we can canonically apply the coadjoint orbits method to any quantum system. For any such system, $G=\cM$ will be manipulated as if it was a matrix group, so that the adjoint transformation $Ad$, which is the natural action of the group on the Lie algebra, is just
\begin{equation}
Ad_{U}(\mathcal{O})=U\mathcal{O}U^{-1}\, ,
\end{equation} 
for $U\in \cM$ and $\cO\in \mathfrak{u}(\cH)$. The dual space $\mathfrak{u}(\cH)^{\ast}$ of the Lie algebra $\mathfrak{u}(\cH)$ can be identified with the set of states, which we generically denote by $\rho$. The pairing $\langle\rho,\mathcal{O}\rangle$ defined by the Killing form between the dual space and the Lie algebra is the usual expectation value, namely,\footnote{For visual simplicity, in what follows we define the expectation value of an operator $\mathcal{O}$ in the state $\rho$ as $\textrm{Tr}(\rho \mathcal{O})$. Of course, this notation is not correct in QFT, where there are no proper traces, and one should write $\rho (\mathcal{\mathcal{O}})$. This is a notational subtlety, but the comments in this section apply to any quantum system, including QFT.}
\begin{equation}
\langle\rho,\mathcal{O}\rangle = \textrm{Tr}(\rho \mathcal{O})\;.
\end{equation}
The coadjoint transformation $Ad^{*}_{U}$ is the natural action of the group on the dual space, and it is defined such that the previous pairing is left invariant.\footnote{In the present notation, if we evolve the state forward in time, we need to evolve the operators backwards for the same amount of time so as to keep the expectation value fixed.} This implies
\begin{equation}
Ad^{*}_{U}(\rho)=U\rho U^{-1}\;.
\end{equation}
For this unitary group $G$, the definition of the Maurer-Cartan form~(\ref{Q}) is equivalent to the definition of the instantaneous gate~(\ref{sch}) in the complexity discussion,
\begin{equation}
H(s)=i \dot U(s) U^{-1}(s)\;.
\end{equation}
Then, the geometric action associated to the unitary group is given by the integral along the affine parameter of the pairing of $\rho(s)$ with $H(s)$, namely
\begin{equation}
I_{\textrm{Geometric}}=\int ds\, \textrm{Tr}\left[\rho (s) H(s)\right]=\int ds\,  \braket{\psi(s)| H(s) |\psi(s)}\, ,
\end{equation}
where $\rho (s)=U(s)\rho_{0} U(s)^{-1}$ and where the second equality holds for pure initial states. Interestingly, this geometric action is intimately related to the complexity measures $\int ds F_{|\braket{H}|}$ and $\int ds F_{\braket{H^2}}$ defined in \req{H2Ja}. It is exactly equal to the first up to the absolute value appearing in $ F_{|\braket{H}|}$, and it agrees with the second whenever the variance for the instantaneous gate is small, namely, whenever $\braket{H^2} \simeq \braket{H}^2$. This happens in many physical applications, in particular in large-$N$ theories and other semiclassical scenarios in which variances are suppressed with respect to averages.

We thus reach the conclusion that the symplectic structure associated to any quantum system gives rise ---through the coadjoint orbits construction--- to a canonical notion of action/distance on the space of unitaries of the system. Such a notion is intimately related to the previously proposed complexity measures $F_{|\braket{H}|}$ and $F_{\braket{H^2}}$. In this sense, besides satisfying all requirements explained in the rest of the paper and giving rise to tight lower bounds for circuit complexity, these measures provide ---at least for systems of small quantum variance--- a realization of the ``complexity equals action'' idea, where ``action'' stands here for the geometric action canonically associated to the quantum system. Interestingly, in 2d CFTs, when doing the pullback to the coadjoint orbits of the Virasoro group, such action is equivalent to Poliakov's two dimensional gravity \cite{Alekseev:1988ce}, as rederived in \cite{Caputa2017} from a quantum complexity perspective using  $ F_{\braket{H^2}}$.

Let us also mention that the symplectic form defined in each orbit of $\mathfrak{g}^*$ ---which is usually called ``Kirillov-Kostant form'' and appears explained in detailed in Appendix~\ref{ap2}--- simply becomes the expectation value of the usual matrix commutator, namely, it exactly agrees with the canonical symplectic form $\Omega$ defined above ---see \req{Ommm}.

As we have mentioned before, in most approaches to QFT complexity, the gate set $\mathcal{G}$ is chosen so that it generates a subgroup of $\mathcal{M}$. In this minisuperspace-like setup, the only elements of $\mathcal{M}$ which can be reached using products of gates are those belonging to such subgroup. In that case, the geometric action functional subject to the appropriate constraints behaves in exactly the same manner as the geometric action of the subgroup generated by the gate set $\mathcal{G}$ without constraints---\ie as in \req{consi}.

\subsection{Quantum action, Hamilton-Jacobi and the semiclassical limit}

As we have seen, the Hilbert space of a quantum system has the structure of a phase space, that is, it is canonically endowed with a symplectic structure. Not only that, as shown \eg in \cite{Ashtekar:1997ud}, Heinseberg's (or Schr\"odinger's) quantum evolution corresponds with the Hamiltonian flow with respect to its canonical symplectic structure and the ``classical'' Hamiltonian, given by $H(\psi)=\langle\psi \vert H\vert\psi\rangle$ at every point $\ket{\psi}\in \cH$ of the phase space (omitting any potential issues regarding the possibility of $H$ being unbounded). As for the Poisson bracket, it is just defined in the conventional way from the canonical symplectic structure.

In more detail, the situation is as follows. Consider a fixed quantum system with associated Hamiltonian $H$ and Hilbert space $\cH$, which we consider as an infinite dimensional K\"ahler manifold $(\cH_{\mathbb{R}},J,\Omega)$. The existence of a symplectic form on $\cH_{\mathbb{R}}$ allows to define the notion of ``Hamiltonian vector field''. More precisely, a vector field $X \in \mathfrak{X}(\cH_{\mathbb{R}})$ is said to be Hamiltonian if there exists a function $f\in C^{\infty}(\cH_{\mathbb{R}})$ such that
\begin{equation}
\iota_X \Omega = df\, .
\end{equation}
If this is the case, we write $X_f = X$. Such function is unique modulo additive constants. Furthermore, the symplectic structure allows to define the notion of ``Poisson bracket'' on $C^{\infty}(\cH_{\mathbb{R}})$, which endows $C^{\infty}$ with the structure of an infinite dimensional Poisson algebra. The Poisson bracket $\left\{ \cdot , \cdot\right\}$ evaluated in two functions $f_1 , f_2 \in C^ {\infty}(\cH_{\mathbb{R}})$ reads:
\begin{equation}
\left\{ f_1 , f_2 \right\} \equiv \Omega(X_{f_1} , X_{f_1})\, .
\end{equation}
Using the given Hamiltonian $H$, we can define a smooth real function $F_H$ on $\cH_{\mathbb{R}}$:\footnote{Or, if $H$ is unbounded, on a dense subspace of $\cH_{\mathbb{R}}$. We will not be concerned about this possibility here.} 
\begin{equation}
F_H\colon \cH_{\mathbb{R}} \to \mathbb{R}\, , \qquad \ket{\psi}\mapsto \bra{\psi}H\ket{\psi}\, .
\end{equation}
Associated to $F_H$ we consider its Hamiltonian vector field, which we denote by $X_H$. The Picard - Lindel\"of's theorem states that, given a state $\ket{\psi_{0}}\in \cH_{\mathbb{R}}$, there exists an interval $I$ with $0\in I$ and a smooth curve $\ket{\psi(s)}\subset \cH_{\mathbb{R}}$ such that $\ket{\psi(0)} = \ket{\psi_0}$ and:
\begin{equation}
\frac{d}{ds}\ket{\psi(s)}\vert_{s_0} = X_H(\ket{\psi(s_0)})\, ,
\end{equation}
which defines the flow associated to $H$ or the ``Hamiltonian flow'' of $F_H$. Now, as shown in \cite{Ashtekar:1997ud}, the quantum evolution of the quantum system as prescribed by Schr\"odinger's equation corresponds with the Hamiltonian flow of $F_H$ as defined above. Therefore, as it happens with any other phase space, we can find local Darboux coordinates $(q_i,p^i)$ for which the symplectic form adopts its standard form. In these coordinates, the equations defining the flow of $X_H$ are the Hamilton equations of motion in their canonical form. As an example, we can consider $\cH = \mathbb{C}^n$ to be a finite dimensional complex Hilbert space equipped with its standard Hermitian metric
\begin{equation}
\sum_{i}dz^i\otimes d\bar{z}^i\, ,
\end{equation}
in canonical complex coordinates $\left\{ z^i\right\}$. Define now the real coordinates
\begin{equation}
q^i \equiv \frac{1}{2}(z^i + \bar{z}^i)\, , \qquad p_i \equiv \frac{1}{2i}(z^i - \bar{z}^i)\, , 
\end{equation}
in terms of which the Riemannian metric $g$ and the symplectic form $\Omega$ read
\begin{equation}
\Omega =\sum_i dq^i \otimes dq^i  + \sum_i dp_i \otimes dp_i\, , \qquad  \Omega =\sum_i dp_i \wedge dq^i\, .
\end{equation}
Hence, as defined above, $(q^i , p_i)$ are indeed Darboux coordinates. Given any smooth function $H$ on $\mathbb{C}^n$ with Hamiltonian vector field $X_H$ we can write
\begin{equation}
X_H = X^i_{H} \frac{\partial}{\partial q^i} + X_{H i} \frac{\partial}{\partial p_i}\, ,
\end{equation}
where $X^i_{H}$ and $X_{H i}$ are local functions depending on both $q^i$ and $p_i$. From the previous equation we obtain
\begin{equation}
\iota_{X_H} \Omega = X_{H i} dq^i - X^i_H dp_i\, ,
\end{equation}
whence equation $\iota_X \Omega = dH$ is locally equivalent to:
\begin{equation}
X_{H i} = \frac{\partial H}{\partial q^i}\, , \qquad X^i_H = - \frac{\partial H}{\partial p_i}\, .
\end{equation}
From the previous equation it follows now directly that the local flow of $X_H$ in the Darboux coordinates $(q^i,p_i)$ yields the standard Hamilton equations. 

The idea that we want to put forward through the rest of the section is that, using Darboux coordinates $(q_i,p^i)$, an explicit action principle can be formulated for each Hamiltonian $H$ on $\cH$, with some intriguing potential applications. The generalized action would take the usual form
\begin{equation}
S_{\rm Quantum}=\int\limits_{t_{i}}^{t_{f}} dt \left[ \sum\limits_{i\in \mathcal{H}} p_{i}\dot{q}_{i}-H(p,q)\right]\, \;,
\end{equation}
which is just constructed so as to obtain the Heisenberg equations of motion. In other words, once we know the instantaneous Hamiltonian and its expectation value (the integrand of the geometric action), the previous equation allows us relate it to the actual action.

This structure has a direct pullback into the appropriate semiclassical phase space of the problem. Calling $\mathcal{H}_{\textrm{class}}$ to such subspace, this is just
\begin{equation}
S_{\rm Quantum}=\int\limits_{t_{i}}^{t_{f}} dt \left[ \sum\limits_{i\in \mathcal{H}_{\textrm{class}}} p_{i}\dot{q}_{i}-H(p,q)\right]\, \;.
\end{equation}
From this formulation it is simple to understand how complexity measures are related to actions. In particular, the geometric action arises by stating that the computational cost is equal to the infinitesimal change in action that arises by a change of the final time, given by
\begin{equation}\label{tf}
\partial S_{\rm Quantum}/\partial t_{f} =-H\, ,
\end{equation}
which is nothing but the Hamilton-Jacobi equation in (not so much) disguise. More generically, one might define a cost as the absolute value of the infinitesimal change in action that arises by a change of the endpoint in all possible directions, including time. This leads to\footnote{This is similar in spirit to the analysis recently performed in \cite{Bernamonti:2019zyy}. It would be interesting to clarify the connection between our approach here and the one in that paper. Note however that an inhomogeneous cost function is used in \cite{Bernamonti:2019zyy}.}
\begin{equation}
d S_{\rm Quantum} = \sum\limits_{i\in \mathcal{H}}p_{i}dq_{i}-Hdt=Ldt\, .
\end{equation}
We thus see that the difference lies in choosing Lagrangians versus Hamiltonians in the definition of the costs. Albeit this ``Hamilton-Jacobi cost'' (Lagrangian) might be more convenient to relate complexity to classical actions, we stress that the geometric action one (the Hamiltonian) is also directly related to the classical action through~(\ref{tf}), but at the same time is more directly connected to the canonical metric in Hilbert space. Besides, we suspect that minimizing the geometric action may be equivalent to minimizing the action whose cost is given by the quantum action integrand (what we just called ``Hamilton-Jacobi cost''), although we do not have conclusive evidence to support this claim.

Hopefully, this formulation makes the connection between complexity and actions mostly trivial. The only non-trivial aspects are the constraints that come from establishing that some tangent vectors of the quantum phase space (the possible Hamiltonians) are allowed while others are not. Apart from that, once we know that a certain Hamiltonian produces a geodesic over a certain time,\footnote{See \cite{Balasubramanian:2019wgd} for recent results in this direction.} the previous equations allow us to connect actions and costs through the expectation value of the instantaneous Hamiltonian. Also, from this formulation it seems clear to us that Fubini-Study-related choices do not play a role in the connection between actions (functionals that actually describe the dynamics of the system) and complexity. Only the geometric action and the ``Hamilton-Jacobi cost'' will do.

\section{Complexity and chaos}\label{chaoss}

The relation between complexity and chaos has attracted some attention recently \cite{Miyaji:2016fse,Brown:2017jil,Magan:2018nmu,universe5040093,Ali:2019zcj,Yang:2019iav}.\footnote{Our approach is very similar to the one in \cite{Miyaji:2016fse}. But here, we are interested in defining chaos as it is conventionally done in classical scenarios, \ie by comparing the evolution of nearby states with the same Hamiltonian. In \cite{Miyaji:2016fse}, however, the states that were compared were evolved with different Hamiltonians. } From our perspective, an important motivation is to provide a complementary, and perhaps more intuitive, approach to quantum chaos and Lyapunov exponents than the one arising from out-of-time-ordered correlators \cite{1969JETP...28.1200L}. In particular, it would be interesting to have a definition of quantum chaos and Lyapunov exponents which reduces to the classical one in the appropriate limits. The objective of this section is to develop a specific proposal in this regard.

Let us start with the most basic and intuitive understanding of chaos, that of extreme sensitivity to initial conditions. In the classical context, it is natural to frame the discussion in phase space. Any point $x$ in a phase space $\mathcal{M}$ is a good initial condition for the classical equations of motion, and in fact there is only one classical trajectory $\gamma_{x}(t)$ going through such a point. Chaotic behavior is said to exist around one point $x$ and along a certain direction $y$ whenever the distance between $\gamma_{x}(t)$ and $\gamma_{x+\delta_{y} x}(t)$  ---whose starting point $x+\delta_{y} x$ is a slight perturbation of $x$ along the direction $y$--- grows exponentially with time. 

There is a small technical problem with this intuitive definition. At a formal level, it is not clear what we mean by ``distance between nearby trajectories". Phase spaces have a canonical symplectic structure built in. 
However, in general, they do not possess a canonical metric. So there might be some ambiguities in statements regarding distance between trajectories. These ambiguities are reminiscent of those  encountered when discussing notions of quantum complexity.

Let us consider the simplest example, consisting of an inverted harmonic oscillator.\footnote{This example has also been considered recently in \cite{Ali:2019zcj} with similar motivations.} This turns out to be mathematically analogous to the case of an infalling particle whose momentum has been perturbed \cite{Magan:2018nmu}, as the solutions to the equations of motions are just boosts, \ie rotations by an imaginary angle. Also, for true chaotic systems, and locally in phase space, this model controls the effective chaotic dynamics near unstable stationary points, as one can observe by expanding the potential around such a point in phase space. The Hamiltonian is given by
\begin{equation}
H=\frac{p^{2}}{2}-\frac{x^{2}}{2}\, .
\end{equation}
The solutions to the equations of motion can be written as
\begin{eqnarray}
q(t)=q_{0}\,\cosh (t)+p_{0}\,\sinh (t)\, , \quad 
p(t)=p_{0}\,\cosh (t)+q_{0}\,\sinh (t)\;.
\end{eqnarray}
Perturbations of the initial conditions $(x_{0},p_{0})\rightarrow (x_{0}+\delta x_{0},p_{0}+\delta p_{0})$ grow similarly as
\begin{eqnarray}
\delta q(t)=\delta q_{0}\,\cosh (t)+\delta p_{0}\,\sinh (t)\, , \quad 
\delta p(t)=\delta p_{0}\,\cosh (t)+\delta q_{0}\,\sinh (t)\;.
\end{eqnarray}
At late times, the two nearby trajectories ``separate exponentially'' in both their position and momentum. But  observe that this separation is a coordinate-dependent statement. As we commented above, a more rigorous invariant statement in terms of a metric in the tangent space to the phase space would be most welcome.

\begin{figure}[t]
\centering 
                \includegraphics[scale=0.45]{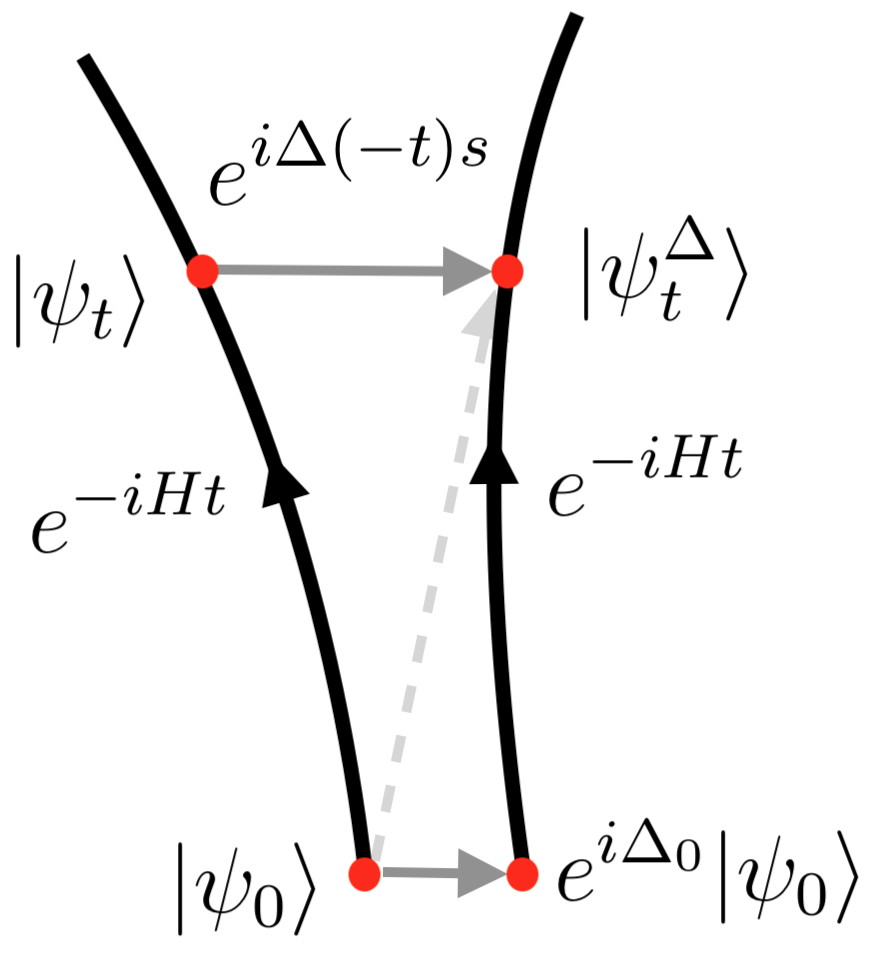} 
        \caption{ \label{figg1}  An initial point in the quantum phase space, $\ket{\psi_0}$, and a nearby perturbed version of it, $e^{i\Delta_0} \ket{\psi_0}$, evolve with physical time $t$ to states $\ket{\psi_t}$ and $\ket{\psi_t^{\Delta}}$ respectively. For each $t$, the complexity between both states measured by some continuous metric $F$ is given, for a sufficiently small perturbation, by $F(\Delta(-t))$, where $\Delta(-t)=e^{-iHt}\Delta_0 e^{iHt}$. The dark gray arrow above corresponds to the straight-line geodesic connecting both states. The pale dashed one corresponds to the one directly connecting $\ket{\psi_0}$ with $\ket{\psi_t^{\Delta}}$ ---see discussion around \req{cfs}.    }
\end{figure}

Before that, let us discuss the quantum mechanical case. Although the system is unbounded and does not have a ground state, the time evolution of operators and states is well defined. Indeed, the solution to Heisenberg's equation reads
\begin{eqnarray}
\hat{q}(t)=\hat{q}_0\,\cosh (t)+\hat{p}_0\,\sinh (t)\, , \quad
\hat{p}(t)=\hat{p}_0\,\cosh (t)+\hat{q}_0\,\sinh (t)\;.
\end{eqnarray}
Let us go back now to the chaotic situation depicted in Fig.~\ref{figg1}, where the initial point in phase space is now a generic quantum state $\vert \psi \rangle$ ---\ie a point in the quantum phase space. Small perturbations are driven by linear combinations of $\hat{p}$, $\hat{q}$, and powers of them. Let us consider the simplest of such perturbations
\begin{equation}
e^{i\Delta_{0}}=e^{i(\hat{p}\delta q_{0}-\hat{q}\delta p_{0})}\;.
\end{equation}
The time evolution of the perturbed state $\vert \psi_{t}^{\Delta}\rangle$ is given by
\begin{equation}
\vert \psi_{t}^{\Delta}\rangle=e^{-iHt}e^{i\Delta_{0}}\vert \psi_{0}\rangle=e^{i\Delta (-t)}\vert \psi_{t}\rangle\;,
\end{equation}
where $\Delta (-t)=e^{-iHt}\Delta_{0}e^{iHt}=\hat{p}(-t)\delta q_{0}-\hat{q}(-t)\delta p_{0}$ is the Heinseberg operator evolved backwards in time. Notice that we can massage such expression so as to obtain
\begin{equation}
\Delta (-t)=\hat{p}\delta q (t)-\hat{q}\delta p (t)\;.
\end{equation}
From this it follows that if we considered a coherent state, $\ket{\psi_t}=\ket{p(t),q(t)}$, then the evolution of the perturbed state would read\footnote{In this relation, the attentive reader will miss a phase in the right hand side of the equation, coming from the non-commutative behavior of $q$ and $p$. We do not include it because it is proportional to the commutator of the two operators, and so it comes at second order in the infinitesimal perturbation $\delta q (t),\delta p (t)$. In any case, this phase does not modify the state as a ray.}
\begin{equation}
\vert \psi_{t}^{\Delta}\rangle=e^{i(\hat{p}\delta q (t)-\hat{q}\delta p (t))}\vert p(t),q(t)\rangle =\vert p(t)+\delta p(t),q(t)+\delta q(t)\rangle\;.
\end{equation}
Now, since the unitary $e^{i\Delta (-t)}$ mediates the transition between the time-evolved unperturbed state $\vert \psi_{t}\rangle$ and the perturbed one $\vert \psi_{t}^{\Delta}\rangle$, its complexity is the relative complexity between both states. So can we compute the complexity of $e^{i\Delta (-t)}$? The answer is surprisingly simple.\footnote{This was not noticed in \cite{Magan:2018nmu}, where the real time $t$ was confused with the protocol ``time'', here denoted by $s$.} As usual, we need to find the shortest geodesic between the identity and $e^{i\Delta (-t)}$ for each $t$. We claim that it is given simply by
\begin{equation}
U_{\textrm{geodesic}}(s)=e^{i\Delta (-t)s}\, ,\quad s\in [0,1]\, .
\end{equation}
This trajectory satisfies the required conditions, namely, it starts at the identity and ends up at the target unitary. Moreover, it is a minimal geodesic because constant Hamiltonians  (in the protocol time $s$) draw minimal geodesics over sufficiently small distances and  $\Delta (-t)=\hat{p}(-t)\delta q_{0}-\hat{q}(-t)\delta p_{0}$ can be made arbitrarily small by letting $\delta q_{0}$ and $\delta p_{0}$ go to zero. This is indeed the conventional double scaling limit used when defining chaos in classical theories.\footnote {See for example Ref. \cite{Book2016}.} Basically, the leading Lyapunov contribution is isolated by taking the limit of infinite time. But, before doing so, one has to take the size of the initial perturbation to zero, so that the linear approximation is valid. 
We conclude that the ``relative complexity'' between the evolved state and the perturbed one is given, in the limit of sufficiently small initial perturbation, by
\begin{equation}
C_{\vert \psi_{t}\rangle\rightarrow \vert \psi_{t}^{\Delta}\rangle}=F(\Delta (-t))\;,
\end{equation}
where $F$ is the complexity measure we choose to use. Interestingly, for all state-dependent costs defined in Section \ref{statedep} it is easy to show that
\begin{equation}
F(\Delta (-t))=F(\Delta_{0})\;,
\end{equation}
so they are insensitive to the ``separation'' of nearby trajectories. 

In fact, it is interesting to notice that such state dependent norms define distances for the classical theory as well, just by the natural pull-back of the appropriate metric to the semiclasical phase space. The fact that ``distance'' does not increase in the classical system either, is rooted in the following equation 
\begin{equation}
H_{\textrm{classical}}=p(t)\delta q(t)-q(t)\delta p(t)=p_{0}\delta q_{0}-q_{0}\delta p_{0}\;,
\end{equation}
\ie although both pairs, ${q(t),p(t)}$ and ${\delta q(t),\delta p(t)}$, show exponential behavior, the previous combination is time independent. But such combination can be used to define distances, such the geometric action one. This neatly shows the sometimes unnoticed dependence of the chaotic analysis in the distance chosen.

This of course does not mean that Lyapunov growth cannot be seen using state dependent measures. The Lyapunov growth just arises by computing the geometric complexity growth from any given fixed point $\vert \psi_0 \rangle$ towards the evolution of the perturbed state $\ket{\psi_t^{\Delta}}$  ---this would correspond to the pale dashed arrow in Fig.~\ref{figg1}. In equations,
\begin{equation}\label{cfs}
C_{\vert \psi_0\rangle\rightarrow \vert \psi_{t}^{\Delta}\rangle}=p_0\delta q(t)-q_0\delta p(t)\;,
\end{equation}
which does grow exponentially fast with a rate given by the Lyapunov exponent, and it is indeed the classical definition of chaos. This was used in \cite{Magan:2018nmu} to obtain the connection with the chaotic behavior in black-hole physics. At any rate, having the instantaneous Hamiltonian $\Delta (-t)$ it is very simple to device norms in phase space that are sensititve to the Lyapunov growht of the infinitesimal perturbations.

To generalize these observations, let us first review  the definition of chaos in classical systems. In phase space $\mathcal{M}$ ---whose points we generically denote by $x$--- we have a Hamiltonian flow which can be written as
\begin{equation}
\dot{x}_{i}(t)=f_{i}(x(t))\;.
\end{equation}
We can expand such equation around any given point $x^*$ obtaining
\begin{equation}
\dot{x}_{i}(t)=f(x^{*})+\sum\limits_{j}\frac{\partial f_{i}(x)}{\partial x_{j}}\vert_{x^{*}}\Delta x_{j}(t)\;.
\end{equation}
Doing the same thing for a nearby trajectory $y(t)=x(t)+\delta x(t)$ we get a linear equation for the difference
\begin{equation}\label{linearized}
\delta \dot{x}_{i} (t)=\sum\limits_{j}\frac{\partial f_i(x)}{\partial x_{j}}\vert_{x^{*}}\delta x_{j} (t)\equiv \sum\limits_{j}L_{ij}\delta x_{j}\;.
\end{equation}
Of course, this linear approximation breaks down after some time. Since to isolate the Lyapunov growth we need to wait a sufficient amount of time, we need to take the limit of very small perturbation before, as commented earlier. The matrix $L_{ij}=\frac{\partial f_i(x)}{\partial x_{j}}\vert_{x^{*}}$ is called the Jacobian matrix, and it determines the stability properties of stationary points in phase space. The linear equation for the deviation is solved by standard methods and we might get oscillatory versus non-oscillatory behavior and dilatation versus contraction behavior ---see \cite{Book2016}.

The quantum complexity story repeats itself in this general scenario. Locally in phase space we can always choose a coordinate chart in which we have $n$ position operators $\hat{q}_{i}$ and $n$ conjugate momentum operators $\hat{p}_{i}$. A generic small perturbation can be parametrized by
\begin{equation}
\Delta_{0}=\sum\limits_{i}\hat{p}_{i}\delta q_{i}^{0}-\hat{q}_{i}\delta p_{i}^{0}\, .
\end{equation}
The unitary transforming $\vert \psi_{t} \rangle$ into $\vert \psi_{t}^{\Delta} \rangle =e^{-iHt}e^{i\Delta}\vert \psi \rangle$ is given by $e^{i\Delta(-t)}$. Most importantly, we again claim that the complexity geodesic for such unitary is
\begin{equation}\label{geo}
U_{\textrm{geodesic}}(s)=e^{i\Delta (-t)s}=e^{i(\sum\limits_{i}\hat{p}_{i}(-t)\delta q_{i}^{0}-\hat{q}_{i}(-t)\delta p_{i}^{0})s}=e^{i(\sum\limits_{i}\hat{p}_{i}\delta q_{i}(t)-\hat{q}_{i}\delta p_{i}(t))s}\;,
\end{equation}
for $s\in [0,1]$, and where $\delta q_{i}(t)$ and $\delta p_{i}(t)$ are the solutions to the linearized equations of motion~(\ref{linearized}). Such linearized equations can refer to some semiclassical phase space or to a bigger quantum phase space. The reason that~(\ref{geo}) is a minimal geodesic is the same as before. For the linearized equations to be valid ---so that a stability analysis and computation of Lyapunov exponents can be performed--- we must take the initial perturbation to zero. Therefore, even if it grows with time, it remains as small as we wish. Since constant Hamiltonians are minimal geodesics for sufficiently small perturbations, the statement follows. Given that we have found the instantaneous Hamiltonian as a function of the Jacobian matrix (from which the Lyapunov spectrum can be computed), the connection between complexity and chaos is transparent.

 Summarizing, we have made the following generic observations:
i) The standard definition of chaos in classical systems relies only on the existence of a symplectic manifold and a Hamiltonian flow on it. Since quantum evolution can be seen as classical dynamics on quantum phase space, we can study quantum chaos using the classical definition on quantum phase space. This approach has the right semiclassical limit by construction.
ii) The instantaneous Hamiltonian can be found in terms of the Jacobian matrix and Lyapunov exponents of the chaotic process. Since we can make the initial perturbation as small as we want, this instantaneous Hamiltonian actually defines a minimal geodesic, whose associated computational cost can be easily found for a given complexity measure.
iii) State dependent costs are insensitive to the Lyapunov growth. But this feature happens both for quantum and classical dynamics! This is a reflection of the fact that Lyapunov growth is a coordinate/metric-dependent statement, as explained for example in \cite{Book2016} . In any case, we remark that the instantaneous Hamiltonian is determined by the solutions to the linearized stability equation, and clearly contains all the information associated to the Lyapunov spectrum.

\section{Discussion} \label{discuss}

There has been much discussion concerning possible QFT complexity measures. The situation is a bit chaotic, sometimes with even more than one proposal per research group. This discussion was partially ignited by arguments \cite{Brown:2017jil} claiming that canonical metrics\footnote{Here the word ``canonical'' refers to a mathematical perspective. As discussed earlier, in Hilbert space it would correspond to the inner product, in projective space it would correspond to the Fubini-Study metric, for a finite system, in the unitary manifold it would be given by the trace, etcetera.} are not good enough for complexity purposes, as those would prevent the existence of large distances in the complexity manifold.

Perhaps the main lesson of the present paper is that canonical metrics can still be used ---and indeed they seem to be the best candidates--- provided one includes constraints in the geodesic action. This conclusion is naturally reached by considering the condition which determines whether or not a given measure indeed provides a lower bound to the quantum complexity of a given set of gates. This analysis, apart from ruling out some possibilities, still leaves us with a huge zoo of putative complexity metrics. It turns out to be convenient to characterize those cost functions by: i) the way the treat gates in the gate set; ii) the way they treat gates outside the gate set. A natural principle to choose within this zoo has been put forward above. Since the characterizing criterion is that valid measures give rise to lower bounds to quantum complexity, we should choose the one providing the tightest lower bound, \ie the one providing the largest distances. This maximization can be accomplished by assigning infinite costs to gates outside the gate set, and by establishing a hierarchy of costs within the gate set. Such infinite penalties can be modeled by constraints in a canonical complexity geometry (which does not assign arbitrary penalties to arbitrary gates).

Regarding the hierarchies that arise by comparing metrics within the gate set, the strongest candidate seems to correspond to the norm of the instantaneous Hamiltonian, $F_{\vert\vert H\vert\vert}$ which, on the other hand, is quite challenging from a technical perspective and somewhat disfavored from a physical one. The second to best metrics are the ones induced by the canonical metrics in Hilbert space and the unitaries manifold, respectively, $F_{\langle H^{2}\rangle}$ (or $F_{|\braket{H}|}$) and $F_{{\rm Tr}H^2}$. These cannot be directly compared, but in order to make sense of the second, one is forced to include a constraint on the total energy of the process in order to prevent violations of Lloyd's bound, while $F_{\langle H^{2}\rangle}$ has the constraint built in.

Including additional arguments into the discussion,
the best choice seems to be $F_{\langle H^{2}\rangle}$. This measure arises from the usual Hermitian metric in Hilbert space  which, along with the associated symplectic form ---given by the expectation value of commutators--- equip every quantum Hilbert space with a K\"ahler structure \cite{Kibble:1978tm,Ashtekar:1997ud}. In a completely natural fashion, this symplectic form can be understood as the  Kirillov-Kostant form associated to the group of unitary transformations acting on such Hilbert space, which makes transparent the connection between the cost function $F_{\langle H^{2}\rangle}$, and the so-called ``coadjoint (or geometric) actions'' \cite{Kiri} . At least for systems of small quantum variance, $F_{\braket{H^2}}$ provides a canonical realization of the ``complexity equals action'' idea ---as previously observed in  \cite{Caputa2017} in the particular context of Virasoro circuits and generalized coherent states.



Exploiting the phase-space structure of Hilbert space, we also explored the possibility of defining a classical action controlling the quantum dynamics.
The purpose of such a formulation ---apart from its inherent interest--- resides in comparing such action with the complexity functionals, elucidating what measures are favored. 
This ``quantum action'' has the right pullbacks to classical actions on the appropriate semiclassical phase spaces. Using such redefinition of a quantum mechanical system, we have seen how the geometric action cost arises and, along the way, we found an alternative ``Lagrangian'' cost function, whose integral over time is by construction the action. 

Finally, having laid out these arguments and structures, we have used them to analyze a particular physical problem ---which has attracted some attention recently \cite{Miyaji:2016fse,Brown:2017jil,Magan:2018nmu,universe5040093,Ali:2019zcj,Yang:2019iav}--- concerning the relation between chaos and complexity. The fact that the symplectic structure extends beyond semiclassical approximations to the full Hilbert space, suggests defining quantum chaos using the conventional classical definition applied to the quantum phase space. This is slightly different from the out-of-time-ordered correlators approach \cite{1969JETP...28.1200L}. 
The challenge  lies here in obtaining the instantaneous Hamiltonian connecting two nearby trajectories as a function of time. This can be accomplished as a function of the usual incidence or stability matrix, whose eigenvalues are basically the Lyapunov exponents. Moreover, such Hamiltonian, in the double limit appropriate for the definition of chaos (long times but infinitesimal perturbations), draws a minimal geodesic. With such input, it is easy to study complexity using the zoo of complexity metrics.


\section*{Acknowledgments}

\noindent We thank Pawel Caputa, Horacio Casini, Rob Myers, Diego Pontello and Joan Simon for useful discussions. The work of PB and JMM was supported by the Simons foundation through the It From Qubit Simons collaboration. The work of CSS was supported by the Humboldt Research Fellowship ESP 1186058 HFST-P from the Alexander Von Humboldt Foundation. CSS would like to thank the members of the theory group at Centro At\'omico Bariloche for their warm hospitality.

\appendix

\section{Physical examples of instantaneous Hamiltonians}\label{ap1}
There are some situations in which \req{nested} can be solved explicitly. The simplest ones correspond to unitary transformations for which the instantaneous Hamiltonian is constant along the path. In that case, $U(s)=e^{-i H s}$, and $H(s)=H $. But there are more interesting examples which can be worked out as well. These include the cases of: commuting generators, piecewise linear paths, protocols generated by symmetry groups ---including the case of Virasoro protocols in $d=2$ CFTs, which we review separately---  and protocols generated by Generalized free fields. 

\subsection{Commuting generators} 
A simple situation in which \req{nested} can be summed corresponds to the case in which $\mathcal{O}(s)$ is a linear combination of mutually commuting operators, namely
\begin{equation}
\mathcal{O}(s)=\sum_I \theta_I(s) K_I\, , \quad \text{where} \quad [K_I,K_J]=0\, , \quad \forall\, I, J\, .
\end{equation}
In that case, the instantaneous Hamiltonian is simply given by 
\begin{equation}
H(s)=\frac{d\mathcal{O}(s)}{ds}=\sum_I \dot \theta_I(s) K_I\, .
\end{equation}
As we said before, it is in general possible that the set of generators is continuous instead of discrete. In that case, we would have something like 
\begin{equation}
\mathcal{O}(s)=\int dk\, \theta_k(s) K(k)\, , \quad \text{where} \quad [K(k),K(k')]=0\, , \quad \forall k, k'\, .
\end{equation}
Similarly to the discrete case, the instantaneous Hamiltonian reads then 
\begin{equation}
H(s)=\frac{d\mathcal{O}(s)}{ds}=\int dk\, \dot\theta_k(s) K(k)\, .
\end{equation}
Examples of instantaneous Hamiltonians of this kind have appear in the complexity literature \eg in \cite{Chapman:2017rqy} in the context of free scalar fields, and in \cite{Magan:2018nmu} connecting boost transformations, chaos and black holes.

\subsection{Infinitesimal paths} 
A very similar situation to the one just described occurs when the unitary $U(s)$ is generated by an infinitesimal Hermitian operator, namely, when
\begin{equation}
U(s)=e^{-i\mathcal{O}(s)}\, , \quad \text{where} \quad \mathcal{O}(s)=\varepsilon \, h(s) \quad \varepsilon\ll 1\, .
\end{equation}
Examples of such infinitesimal paths have been consider \eg in \cite{Belin:2018bpg} in the context of conformal deformations of the vacuum state in two-dimensional CFTs. In this case, all commutators in \req{nested} are order $\varepsilon^2$ or higher. Hence, at leading order in $\varepsilon$ the situation is identical to the one considered in the previous subsection, and we trivially have
\begin{equation}
H(s)=\varepsilon \frac{dh(s)}{ds} +\mathcal{O}(\epsilon^2)\, .
\end{equation}

\subsection{Piecewise linear paths}\label{piece}
Imagine now that we can construct $U_f$ as a finite sequence of ``small unitaries'' of the form
\begin{equation}\label{usis}
U_f=e^{-i h_{(N)}}e^{-i h_{(N-1)}}\cdots e^{-i h_{(1)}}\mathds{1}\, .
\end{equation}
This is the usual setup in quantum computation problems. In that context, the small unitaries would correspond to the gates of the circuit. From the point of view of the unitaries manifold  $\mathcal{M}$, \req{usis} defines a discrete sequence of points: $U_0\equiv \mathds{1}$, $U_1\equiv e^{-i h_{(1)}}\mathds{1}$, $U_2\equiv e^{-i h_{(2)}}e^{-i h_{(1)}}\mathds{1}$, and so on. We can likewise define a curve $U(s)$ consisting of straight  lines in $\mathcal{M}$ connecting consecutive intermediate unitaries for that sequence. For values of the affine parameter in the range $(j-1)/N < s <j/N$, ($j \in \mathbb{N}$), we have
\begin{equation}\label{usis2}
U(s)=e^{-iN \left(s-\frac{j-1}{N} \right) h_{(j)}} e^{-i h_{(j-1)}} \cdots e^{-i h_{(1)}} \mathds{1}\, .
\end{equation}
After applying the unitary $e^{-i Ns h_{(1)}}$, which approaches $e^{-ih_{(1)}}$ as $s\rightarrow 1/N$ on $\mathds{1}$, we introduce a new unitary which acts on $e^{-i h_{(1)}} \mathds{1}$ and approaches $e^{-ih_{(2)}}$ as $s\rightarrow 2/N$, and so on. With this definition, $U(s)$ is a continuous curve which is not differentiable in general at the points $s=j/N$. It is not difficult to see that the instantaneous Hamiltonian is given by\footnote{Consider for example the infinitesimal evolution from $s=(j-1)/N$ to $s=(j-1)/N+ds$. Then, we have from the definition of instantaneous Hamiltonian \req{usds}: $U((j-1)/N+ds)=e^{-iH((j-1)/N) ds} U((j-1)/N)$. Comparing with \req{usis2}, it follows that $H(s)=N h_{(j)}$ in that interval.}
\begin{equation}
H(s)=N h_{(j)} \quad \text{for} \quad (j-1)/N < s <j/N\, .
\end{equation}
The $s$-dependence only appears through sudden jumps in $H(s)$ as $Ns$ takes integer values. Piecewise protocols play an important role in Nielsen's original proposal for approximating complexity by the length of continuous paths on $\mathcal{M}$ \cite{2005quant.ph..2070N}.  We use this setup in Section \ref{bounds}.

\subsection{Symmetry transformations} \label{symmm}
Naturally, the nontriviality of \req{nested} also disappears when the sequence of nested commutators can be  performed explicitly, and the series resummed. A prototypical case occurs when the path in the unitaries manifold can be thought of as parametrizing a continuous set of elements of some symmetry group $G$. This has been emphasized in several papers which try to provide notions of complexity valid for quantum field theories \cite{Hackl:2018ptj,Magan:2018nmu,Guo:2018kzl,Caputa:2018kdj,Chapman:2018hou}, and it has been more or less implicitly exploited in many other related works ---see \eg \cite{Yang:2017nfn,Kim:2017qrq,Jiang:2018nzg,Jefferson:2017sdb,Hashimoto:2017fga,Khan:2018rzm,Alves:2018qfv,Camargo:2018eof,Ge:2019mjt}.

 In this context, a finite sequence of gates of the form \req{usis} would read
\begin{equation}
U_f=U_{g_N}U_{g_{N-1}}\cdots U_{g_1}\mathds{1}=U_{g_N g_{N-1}\cdots g_1}\, ,
\end{equation}
where $g_{i}\in G$ $\forall i=1,\dots, N$, and $U_{g_j}$ is some representation of $G$ acting on the Hilbert space of the system. In the continuous case, $U(g(s))$ represents a path in $G$, and the instantaneous Hamiltonian is represented by some element of the Lie algebra $\mathfrak{g}$. In particular, \req{u0} can now be written in terms of the group elements as\footnote{Observe that while we write generic unitary operators as $U=e^{-i\mathcal{O}}$, \ie with an explicit factor of $(-i)$, whenever $U$ is the representation of a symmetry group, we follow the more standard mathematical notation and consider the exponential map $g=e^{\mathcal{O}}$ without such factor. Then, one must remove the $(-i)^j$ factor from \req{nested} when computing $Q(s)$ in that case. }
\begin{equation}
g(s+ds)=e^{Q(s) ds}\circ g(s)\, ,
\end{equation}
where $\circ$ stands for the group multiplication, and $Q(s)\in \mathfrak{g}$ plays the role of the instantaneous Hamiltonian. Then, the solution to the group-theoretical version of the Schr\"odinger equation \req{sch} reads
\begin{equation}\label{Qs}
Q(s)=\left.\frac{d}{ds'}(g(s')\circ g^{-1}(s))\right|_{s'=s}\, ,
\end{equation}
which is the adjoint transformation of the Maurer-Cartan form ---see Section \ref{geomA} below. Equivalently, $Q(s)$ can be also computed using the infinite sum of nested commutators in \req{nested} by writing the unitary representations of the group as exponentials of elements of the Lie algebra, \ie $U(g(s))=e^{\mathcal{O}(s)}$, with $\mathcal{O}(s)=\theta_a(s) T_a$, where the $\theta_a(s)$, are certain functions of the affine parameter, and $T_a$ are the generators of $\mathfrak{g}$. 

Consider for example the case of the Heisenberg group. We can parametrize a generic element of the Lie algebra as $\mathcal{O}(s)=x(s) X+ y(s) Y+z(s) Z$, where the only non-vanishing commutator reads $[X,Y]=Z$. Then, using \req{nested}, we are left with $Q(s)=\dot{x}(s) X+\dot{y}(s) Y+\dot{z}(s) Z+\frac{1}{2!} [\mathcal{O}(s),\dot{\mathcal{O}}(s)]$ plus higher order commutators. It is straightforward to obtain $[\mathcal{O}(s),\dot{\mathcal{O}}(s)]=(x(s) \dot y(s)- \dot x(s) y(s))Z$, and since $Z$ belongs to the center of $\mathfrak{g}$, all such commutators vanish. Therefore, we are left with
\begin{equation}\label{Qss}
Q(s)=\dot{x}(s) X+\dot{y}(s) Y+\left[\dot{z}(s)+\frac{1}{2}(x(s) \dot y(s)- \dot x(s) y(s)) \right] Z\, .
\end{equation}
Alternatively, we can start from an element of the group and apply \req{Qs}. The appropriate parametrization of $g(s)$ reads 
\begin{equation}
U(g(s))=\begin{bmatrix}
1 & x(s) & z(s)+\frac{1}{2} x(s) y(s) \\
0 & 1 & y(s) \\
0 & 0 & 1
 \end{bmatrix} \, ,
 \end{equation}
where the nonvanishing components of the Lie algebra generators read $X_{12}=Y_{23}=Z_{13}=1$. Then,  applying $Q(s)=dg(s)/ds\circ g^{-1}(s)$ we are left precisely with \req{Qss}.

\subsection{Virasoro protocols} 
Another particularly relevant example belonging to the class described in the previous subsection corresponds, in the context of conformal field theories, to Hermitian operators $\mathcal{O}(s)$ constructed from the energy-momentum tensor. In \cite{Caputa:2018kdj}, these were proposed to play a crucial role in the connection between complexity and gravity.
While \req{nested} cannot be resummed for generic CFTs, the situation changes in two dimensions, where such operators generate the Virasoro algebra.

In the framework described in the previous section, the group $G$ is the group of diffeomorphisms of the circle $f(\sigma)$, with the group product given by composition of functions $(f\circ g )(\sigma)=f(g(\sigma))$. An infinitesimal group transformation is now an infinitesimal diffeomorphism, which can be written as $\sigma\rightarrow \sigma+\epsilon(\sigma)$. Therefore, in abstract group terms, the global and complexity frames are related as:
\begin{equation}
f(s+ds,\sigma)=e^{-iH(s)}\circ f(s,\sigma)\nonumber\;,
\end{equation}
where
\begin{equation}\label{gateV}
H(s)\equiv i\int^{2\pi}_0\frac{d\sigma}{2\pi}\epsilon(s,\sigma)  T(\sigma)= i\sum_{n\in \mathbb{Z}}\epsilon_n(s) \left(L_{-n}-\frac{c}{24}\delta_{n,0}\right)\, ,
\end{equation}
is an element of the Virasoro algebra, $L_{n}$ its generators, and we have the usual expansions
on the cylinder of size $2\pi$
\begin{eqnarray}
T(\sigma)=\sum_{n\in \mathbb{Z}} \left(L_n-\frac{c}{24}\delta_{n,0}\right)e^{ -i n \sigma}\, ,\, \quad
\epsilon(\tau,\sigma)=\sum_{n\in \mathbb{Z}} \epsilon_n(\tau)e^{ -i n \sigma}\, .
\end{eqnarray}
We can invert the previous relations to obtain
\begin{eqnarray}
L_n=\int^{2\pi}_0\frac{d\sigma}{2\pi}T(\sigma)e^{in\sigma}+\frac{c}{24}\delta_{n,0},\, \quad
\epsilon_n(\tau)=\int^{2\pi}_0\frac{d\sigma}{2\pi}\epsilon(\tau,\sigma)e^{in\sigma}\, .
\end{eqnarray}
In this context, the instantaneous gate equation can be solved. By expanding the infinitesimal gate near the identity we get
\begin{equation}
f(s+ds,\sigma)=(1+\epsilon(s,\sigma)ds)\circ f(s,\sigma)\nonumber
\end{equation}
Composing the diffeomorphisms in the r.h.s. and expanding the l.h.s. we arrive to:
\begin{equation}
\partial_s f(s,\sigma)ds=\epsilon(s,\sigma)\circ f(s,\sigma)ds=\epsilon(s,f(s,\sigma))ds \nonumber
\end{equation}
or equivalently
\begin{equation}
\epsilon(s,\sigma)=\partial_s f(s,F(s,\sigma))=-\frac{\partial_s F(s,\sigma)}{\partial_\sigma F(s,\sigma)}\nonumber
\end{equation}
where we introduced the inverse function $F(s,f(s,\sigma))=\sigma$.

So, given a time dependent conformal transformation $f(s,\sigma)$ (the output of this set of protocols), we can find the instantaneous Hamiltonian at time $s$, defined by $\epsilon(s,\sigma)$. As in the previous section, we just need to represent such unitary in the appropriate Hilbert space and we would be done. We refer to \cite{Caputa:2018kdj} for more details on this setup as well as for holographic applications.

\subsection{Generalized free fields} 
Another interesting case, which has not been considered in the literature before--- corresponds to the so-called Generalized free fields \cite{ElShowk:2011ag}. A free field is often defined as one obeying a linear equation of motion ---\eg the Klein-Gordon equation for a free scalar. In such scenario, new solutions of the equations of motion can be constructed as superpositions of others, and one can arrive at a non-interacting freely-generated Fock space of excitations.

One may wonder if both statements are equivalent, namely, if the fact that the field obeys a linear equation of motion and the fact that it displays a non-interacting freely-generated Fock space of excitations imply one another. This is not the case. A Fock space of excitations can be obtained whenever correlation functions factorize into products of two point functions, \ie when
\begin{equation}
\textrm{Tr}(\rho\, \mathcal{O}(x_{1})\cdots \mathcal{O}(x_{2n}))=\frac{1}{2^{n}}\sum\limits_{\pi}\textrm{Tr}(\rho\, \mathcal{O}(x_{\pi_{1}}) \mathcal{O}(x_{\pi_{2}}))\cdots \textrm{Tr}(\rho \,\mathcal{O}(x_{\pi_{2n-1}}) \mathcal{O}(x_{\pi_{2n}}))\;,
\end{equation}
where $\rho$ is a generic state. Fields satisfying this property but not obeying any linear equation of motion are called Generalized free fields, and play a prominent role in the holographic context ---see \eg \cite{Papadodimas:2012aq}.

For us, what is important is the implication of the previous relation on the structure of commutators. In particular, it is easy to see that the previous factorization implies
\begin{equation}
\textrm{Tr}(\rho\, \mathcal{O}(x_{1})\cdots \mathcal{O}(x_{2n})[\mathcal{O}(x),\mathcal{O}(y)])=\textrm{Tr}(\rho\, \mathcal{O}(x_{1})\cdots \mathcal{O}(x_{2n}))\textrm{Tr}(\rho\,[\mathcal{O}(x),\mathcal{O}(y)])\;,
\end{equation}
and in fact it does not depend on the position of the commutator in the string. One concludes that ---at least when inserted into correlation functions--- the commutator of two operators  of this kind is proportional to the identity:
\begin{equation}
[\mathcal{O}(x),\mathcal{O}(y)]_{\rho}=\textrm{Tr}(\rho\,[\mathcal{O}(x),\mathcal{O}(y)])\mathds{1}\;.
\end{equation}
The proportionality factor is state-dependent, but otherwise the commutator is a c-number. Indeed, this is another equivalent way of defining a Generalized free field.

This feature can be used to compute the instantaneous gate, at least for computational costs defined in terms of expectation values ---see Section \ref{bounds}. Indeed, since the commutator is proportional to the identity, the sum~(\ref{nested}) colapses to the first two terms, so that:
\begin{equation}
H_{\textrm{GFF}}(s)=\frac{d\mathcal{O}(s)}{ds}-\frac{i}{2}\left[\mathcal{O}(s),\frac{d\mathcal{O}(s)}{ds}\right]\;.
\end{equation}



\section{Geometric actions}\label{ap2}

The concept of geometric action was developed by Kirillov \cite{Kiri}. It emerged as a byproduct of the development of a new framework to find and classify irreducible representations in group theory. This new framework was called the ``coadjoint orbit method'', since its key ingredient is a generic orbit of the coadjoint representation of the given group $G$. A nice review on the subject, with lots of physics applications, can be found in \cite{Oblak:2016eij}. The present appendix intends to be a small self-contained review. We will not worry about any potential issues associated with $G$ being infinite dimensional.

Given a Lie group $G$ we denote by $\mathfrak{g}$ its Lie algebra. The adjoint action of $G$ on $\mathfrak{g}$ is defined in the usual way as follows
\begin{equation}\label{adjoint}
Ad_{g}(Q)=\frac{d}{ds}(g\circ e^{s Q} \circ g^{-1})\vert_{s =0}\, ,
\end{equation}
and yields a homomorphism $Ad\colon G\to \mathrm{Aut}(\mathfrak{g})$ given by $g\mapsto Ad_g$. We will denote by $Q\in \mathfrak{g}$ the elements of the Lie Algebra, since they generically correspond to charges in physical applications. Also, above and in the following, we denote by $\circ$ the group product. Notice that, in general,  we cannot  directly introduce  the derivative in the parenthesis  without evaluating the group product first. This can only be done when $G$ is a matrix group, which is not the case in general. 
 
Denote by $\mathfrak{g}^{*}$ the vector space dual to $\mathfrak{g}$, that is, $\mathfrak{g}^{\ast}$ the vector space of linear maps from $\mathfrak{g}$ to $\mathbb{R}$. We will denote elements of $\mathfrak{g}^{*}$ by $\psi$, since they will typically correspond to quantum states in physical applications. The pairing between the dual space and the Lie algebra will be denoted by $\langle \psi , Q\rangle $. Using such pairing we can introduce a representation of $G$ on $\mathfrak{g}^{\ast}$ canonically induced by adjoint representation of $G$ on $\mathfrak{g}$. This representation is usually called the ``coadjoint representation'', denoted by $Ad^{\ast}\colon G\to \mathfrak{g}^{\ast}$, and it is defined as follows,
\begin{equation}\label{coadjoint}
\langle Ad^{*}_{g}(\psi), Q\rangle \equiv \langle \psi, Ad_{g^{-1}}(Q)\rangle \;.
\end{equation}
for every $g\in G$ and $Q\in \mathfrak{g}$. Note that the adjoint and coadjoint representations are inequivalent in general. They are equivalent if and only if there exists an equivariant map from the Lie algebra $\mathfrak{g}$ to its dual $\mathfrak{g}^{\ast}$, which is in turn equivalent to the existence of a non-degenerate an adjoint-invariant bilinear form on $\mathfrak{g}$. The coadjoint action of $g\in G$ on an element of $\mathfrak{g}^{*}$ paired with some $Q\in \mathfrak{g}$ is therefore defined as the pairing of that element with the element of $\mathfrak{g}$ resulting from the application of the adjoint action of  $g^{-1}$ on $Q$.
We denote as follows:
\begin{equation}
\mathcal{O}_{\psi}\equiv \lbrace\psi' \in \mathfrak{g}^{*}\,\vert\, \exists \,g\in G\rightarrow  Ad^{*}_{g}(\psi)=\psi'\rbrace\, ,
\end{equation}
the orbit of the coadjoint action on $\mathfrak{g}^{\ast}$ passing through $\psi$. This is defined as the set of all elements of $\mathfrak{g}^{*}$ which can be reached from $\psi$ by applying the coadjoint action of the group. 

The remarkable result proved by Kirillov and Konstant is that every such coajoint orbit can be equipped with a canonical and invariant symplectic form. In order to define this symplectic form on $\mathcal{O}_{\psi}$, we recall that $\mathcal{O}_{\psi}\looparrowright \mathfrak{g}^{\ast}$ is not an abstract manifold but an immersed submanifold of $\mathfrak{g}^{\ast}$. Furthermore, from the very definition of orbit, $G$ acts transitively on $\mathcal{O}_{\psi}$ and hence we have a well-defined smooth submersion,
\begin{equation}
\Theta_{\psi}\colon G \to \cO_{\psi}\, , \qquad g\mapsto Ad^{\ast}_g(\psi)\, .
\end{equation}
Its differential at the identity element yields a linear map of vector spaces,
\begin{equation}
d_{\rm Id} \Theta_{\psi}\colon \mathfrak{g}\to T_{\psi}\cO_{\psi}\, .
\end{equation}
In particular, $T_{\psi}\cO_{\psi} = \mathrm{Im}(d_{\rm Id} \Theta_{\psi})$. Note that $d_{\rm Id} \Theta_{\psi}$ may have kernel, which is given by the Lie algebra $\mathfrak{g}_{\psi}$ of the stabilizer of $\psi$ in $G$. In particular:
\begin{equation}
T_{\psi}\cO_{\psi} =  \mathfrak{g}/\mathfrak{g}_{\psi}\, .
\end{equation} 
With these provisos in mind, Kirillov's symplectic form $\omega$ on $\cO_{\psi}$ at $\psi$ is defined as follows
\begin{equation}
\omega_{\psi}(x_1,x_2) \equiv  \langle \psi, [Q_1,Q_2]\rangle\, ,
\end{equation}
where $Q_1 , Q_2 \in \mathfrak{g}$ satisfy $d_{\rm Id} \Theta_{\psi}(Q_a) = x_a$, $a=1,2$. Clearly, $Q_1$ and $Q_2$ will not be unique in general, whence we need to verify that the definition does not depend on the pre-images of $x_1$ and $x_2$ chosen. Any other pre-image $Q^{\prime}$ can be written as follows:
\begin{equation}
Q^{\prime}_a = Q_a + q_a\, , 
\end{equation}
where $q_a\in \mathfrak{g}_{\psi}$. A direct computation shows that:
\begin{equation}
0 = \frac{d}{dt}\langle Ad^{\ast}_{e{-t q}} \psi, Q\rangle\vert_{t= 0} = \frac{d}{dt}\langle \psi , Ad_{e{t q}}Q\rangle\vert_{t= 0} = \langle \psi , [q,Q]\rangle = \omega_{\psi}(q,Q)\, ,
\end{equation}
for every $q\in \mathfrak{g}_{\psi}$ and $Q\in \mathfrak{g}$. Hence:
\begin{equation}
\langle\psi, [Q_1 + q_1,Q_2 + q_2]\rangle = \langle \psi, [Q_1,Q_2]\rangle + \langle \psi, [q_1,Q_2]\rangle +\langle\psi, [Q_1,q_2]\rangle + \langle \psi, [q_1,q_2]\rangle = \langle \psi, [Q_1, Q_2]\rangle\, ,
\end{equation}
whence $\omega$ is indeed well-defined. The fact that $\omega$ is closed follows from carefully applying the standard formula for the exterior derivative in terms of the Lie bracket and from using the Jacobi identity and the fact that $G$ acts transitively on $\cO_{\psi}$. To prove now that $\omega$ is non-degenerate, we need to show that, for every non-zero $x\in T_{\Psi}\cO_{\psi}$ there exists another element $x^{\prime}\in T_{\Psi}\cO_{\psi}$ such that:
\begin{equation}
\omega_{\psi}(x,x^{\prime}) = \langle \psi , [Q,Q^{\prime}]\rangle\neq 0\, .
\end{equation}
In order to see that this is the case, we proceed by contradiction. Assume that such $x^{\prime}$ does not exist. Then:
\begin{equation}
\langle \psi , [Q,Q^{\prime}]\rangle = 0\, , \qquad \forall \,\, Q^{\prime} \in \mathfrak{g}\, .
\end{equation}
This implies:
\begin{equation}
 0 = \langle \psi , [Q,Q^{\prime}]\rangle = -\frac{d}{dt}\langle \psi , Ad_{e{t  Q}}Q^{\prime}\rangle\vert_{t= 0} = -\frac{d}{dt}\langle Ad^{\ast}_{e{ -t  Q}}\psi , Q^{\prime}\rangle\vert_{t= 0} = 0\, , \quad \forall\,\, Q^{\prime}\in \mathfrak{g}\, .
\end{equation}
Hence:
\begin{equation}
Ad^{\ast}_{e{ -t  Q}}\psi = 0\, ,
\end{equation}
implying $Q \in \mathfrak{g}_{\psi}$, in contradiction with the fact that $x\neq 0$ by assumption. To summarize: every orbit of the coadjoint action of $G$ admits a canonical symplectic form, which in addition can be shown to be invariant under the action of $G$ on the orbit. Hence, every such orbit equipped with its canonical symplectic form can be understood as the phase space of a physical system. In these symplectic manifolds we can explore two avenues:

\begin{enumerate}
	\item Define and study classical systems in such phase spaces. This has the advantage that such systems will feature the $G$ group as a global symmetry, and the group $H$ as a gauge symmetry, a fact that we can use to simplify the corresponding physical system, for example by exploiting the associated conserved currents in the case $G$ acts through Hamiltonian symplectomorphisms.
	
	\item We can try to geometrically quantize the phase space by implementing the different recipes developed in geometric quantization. If successful, this would realize the Hilbert space of the associated quantum system as certain subspace of the space of section of a Hermitian line bundle over the orbit, on which the group $G$ acts through unitary transformations, obtaining thus a unitary representation of the group.
\end{enumerate}

For the complexity discussion, the interesting avenue is the first one, since in general we will work with cases for which we know already the appropriate Hilbert space representation. Therefore, what kind of actions can we define naturally within this framework? The most natural way to proceed is to consider a particle moving on a configuration space with associated symplectic phase space given by an orbit of the coadjoint action equipped with its Kirillov-Konstant symplectic form. Each point $\psi$ in the orbit represents a state, and time evolution draws a trajectory $\psi (s)$ in this phase space. Before establishing what kind of action will control $\psi (s)$, let us see how to parametrize such trajectory. To such end, consider $\psi_{0}$ as the initial point. Since the group $G$ acts transitively on the orbit, we can parametrize the trajectory $\psi (s)$ with a path $g(s)$ in the group $G$, by means of:
\begin{equation}
\psi (s)=Ad_{g(s)}^{*}(\psi_{0})\, .
\end{equation}
Due to the fact that $\psi_0$ may have a non-trivial stabilizer the path $g(s)$ is not unique. This will not cause any problems, since we just have to choose one. One can call such coordinate frame $g(s)$ the ``physical frame''. There is another coordinate frame, which completely defines the trajectory as well. It provides the Lie algebra element $Q(s)$ implementing the transition from $s$ to $s+ds$. This is the instantaneous gate in this framework and it is defined group theoretically as:
\begin{equation}\label{Q}
Q (s)= \frac{d}{d\tau}(g (\tau)\circ g^{-1}(\tau) )\vert_{\tau=s}\;.
\end{equation}
We will denote this as the ``complexity frame'', since it is the one that enters directly in the definition of the metric, as we will see below.

Just as a remark, an important cousin of $Q(s)$ is its adjoint transformation under $g(s)$. This is is called the Maurer-Cartan form:
\begin{equation}
\tilde{Q} (s)=\frac{d}{d\tau}( g^{-1}(\tau) \circ g (\tau))\vert_{\tau=s}\;.
\end{equation} 
We are now ready to define the geometric action. Given the trajectory $\psi (t)$, parametrized in such two different ways, its geometric action is:
\begin{equation}
I_{\textrm{Geometric}}= \int ds \,\langle Ad^{*}_{g(s)} (\psi_{0}) , Q(s)\rangle = \int ds \,\langle \psi_{0} , \tilde{Q}(s)\rangle \;.
\end{equation}

\bibliographystyle{JHEP}
\bibliography{Gravities2}

\end{document}